\documentclass[aps,prd,reprint,superscriptaddress,floatfix,nofootinbib,preprintnumbers]{revtex4-1}

\usepackage[utf8x]{inputenc}
\usepackage{microtype}

\usepackage{graphicx}
\usepackage[dvipsnames]{xcolor}
\usepackage{rotating}
\usepackage[caption=false,singlelinecheck=false]{subfig}
\usepackage{amsmath}
\usepackage{amssymb}
\usepackage{bbold}
\usepackage{amsfonts,dsfont}
\usepackage{bm}
\usepackage{physics}
\usepackage{mathtools}
\usepackage{slashed}
\usepackage{nccmath}
\usepackage{verbatim}
\usepackage{float}

\usepackage{tabu}
\usepackage{booktabs}
\usepackage{multirow}
\usepackage{array}
\usepackage{enumitem}

\usepackage{url}
\usepackage{xspace}
\usepackage{siunitx}
\usepackage{xfrac}
\usepackage{hyperref}
\usepackage[nameinlink]{cleveref}
\usepackage{appendix}
\usepackage{xifthen}
\usepackage{xcolor}
\hypersetup{
	colorlinks,
	linkcolor={blue!75!black},
	citecolor={blue!75!black},
	urlcolor={blue!75!black}
}

\usepackage[normalem]{ulem}

\newcommand{\orcid}[1]{\href{https://orcid.org/#1}{\includegraphics[height=1.7ex,width=1.7ex]{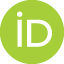}}}

\begin{document}

\title{Towards two-to-two scattering of scalars in asymptotically safe quantum gravity}

\author{Angelo~P.~Chiesa~\orcid{0009-0003-1161-9693}}
\affiliation{Department of Physics and Astronomy, University of Sussex, Brighton, BN1 9QH, U.K.}
\affiliation{Institut f{\"u}r Theoretische Physik, Universit{\"a}t Heidelberg, Philosophenweg 16, 69120 Heidelberg, Germany}

\author{Jan M. Pawlowski~\orcid{0000-0003-0003-7180}}
\affiliation{Institut f{\"u}r Theoretische Physik, Universit{\"a}t Heidelberg, Philosophenweg 16, 69120 Heidelberg, Germany}
\affiliation{ExtreMe Matter Institute EMMI, GSI Helmholtzzentrum f{\"u}r Schwerionenforschung mbH, Planckstr. 1, 64291 Darmstadt, Germany}

\author{Manuel~Reichert~\orcid{0000-0003-0736-5726}}
\affiliation{Department of Physics and Astronomy, University of Sussex, Brighton, BN1 9QH, U.K.}

\begin{abstract} 
We compute the graviton-mediated two-to-two scattering amplitude and cross section for scalar particles in asymptotically safe quantum gravity. Specifically, we compute the full momentum dependence of the scalar-graviton three-point scattering vertex for spacelike momenta with the functional renormalisation group. We also discuss the analytic continuation to the Minkowski branch, and in particular its angular dependence. Then, the timelike part of the vertex is reconstructed and used to compute the scattering amplitude and cross-section. We show that the cross-section reduces to that in General Relativity at small energies, and it respects unitarity in the UV.
\end{abstract}

\maketitle

\section{Introduction} 
\label{sec:intro}

Asymptotically safe quantum gravity \cite{Weinberg:1980gg, Reuter:1996cp} is a promising candidate for a consistent quantum theory of gravity. Within the functional renormalisation group (fRG) framework, substantial evidence has been accumulated for the existence of a non-trivial ultraviolet (UV) fixed point -- the Reuter fixed point -- across increasingly sophisticated truncations. For recent reviews, see \cite{Bambi:2023jiz, Bonanno:2020bil, Dupuis:2020fhh, Knorr:2022dsx, Eichhorn:2022gku, Morris:2022btf, Wetterich:2022ncl, Martini:2022sll, Saueressig:2023irs, Pawlowski:2023gym, Platania:2023srt, Bonanno:2024xne, Reichert:2020mja, Basile:2024oms}.

A central open question concerns the unitarity of asymptotically safe quantum gravity. Addressing this issue requires the computation of physical observables, such as scattering amplitudes on Lorentzian backgrounds. This in turn demands access to momentum-dependent correlation functions in real time. Recent progress has initiated this programme using two complementary strategies. One approach reconstructs Lorentzian correlation functions from Euclidean signature \cite{Bonanno:2021squ, Pastor-Gutierrez:2024sbt, Knorr:2026vax}. This method builds upon the extensive experience with Euclidean fRG computations, but relies on assumptions in the reconstruction of real-time quantities. The second strategy performs calculations directly in Lorentzian signature \cite{Fehre:2021eob, Kher:2025rve, Pawlowski:2025etp, Assant2026}. These computations require specially designed regulators that preserve the causal structure of the theory \cite{Fehre:2021eob, Braun:2022mgx}. Both approaches have successfully been applied to compute the full graviton propagator \cite{Bonanno:2021squ, Fehre:2021eob, Pawlowski:2025etp, Assant2026}. First results for scattering processes have also been obtained, including $e^+e^-\to \mu^+\mu^-$ scattering using a momentum-symmetric approximation of the scattering vertex \cite{Pastor-Gutierrez:2024sbt}, and leading-order two-to-two scattering of distinct scalar fields \cite{Knorr:2026vax}.

\begin{figure*}[t]
    \includegraphics[width=0.6\textwidth]{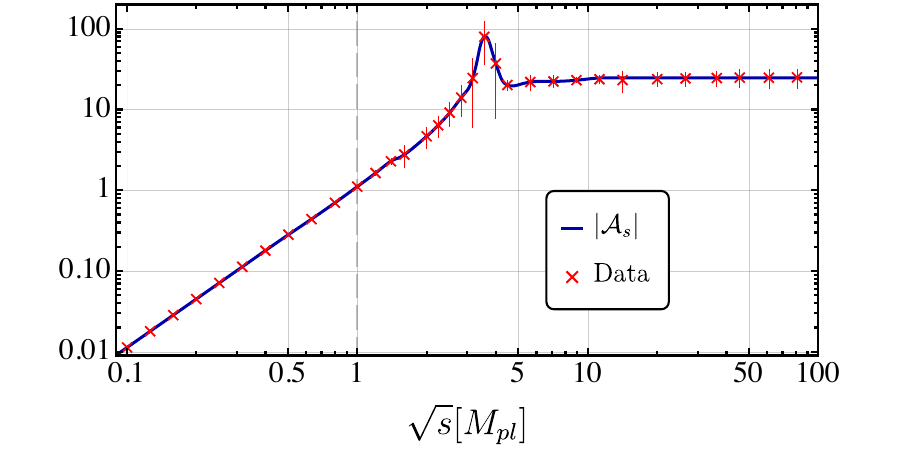}
    \caption{We display the non-perturbative scattering amplitude for graviton-mediated $s$-channel $\phi\phi\to\phi\phi$ scattering, see \Cref{fig:scattering}. The computed data is displayed in red alongside a smooth interpolation in blue for visualisation purposes. The amplitude is shown as a function of the centre of mass energy $\sqrt{s}$, and we display the classical Planck scale with the vertical dashed line. $\mathcal{A}_s$ is bounded for trans-Planckian energies, thus compatible with unitarity.
    }
    \label{fig:AmplitudeResult}
\end{figure*}

In the present work, we present the first computation of a fully momentum-dependent 1PI vertex in Euclidean signature and the reconstruction thereof to Lorentzian signature. Specifically, we compute the two-scalar–one-graviton vertex, which determines the single-graviton exchange contribution to scalar two-to-two scattering. Using this result, we obtain the corresponding scattering amplitude, shown in \Cref{fig:AmplitudeResult}. The amplitude approaches a constant in the UV, satisfies the Froissart bound \cite{Froissart:1961ux}, and is therefore compatible with unitarity.

This paper is organised as follows. In \Cref{sec:AsymptoticallySafeQuantumGravity}, we briefly review scalar scattering in quantum gravity. In \Cref{sec:EuclideanQuantum gravity}, we introduce the functional renormalisation group framework used in this work. The results for the momentum-dependent vertex $\Gamma^{\phi\phi h}$ are presented in \Cref{sec:ResultsForG_hphiphi}. In \Cref{sec:Scalar-gravitonVertexAmplitude}, we use this vertex to construct the Euclidean scattering amplitude. The analytic continuation to Lorentzian signature is discussed in \Cref{sec:LorentzianAmplitude}. Finally, we summarise and discuss our results in \Cref{sec:conclusions}.

\section{Scattering in Quantum Gravity} \label{sec:AsymptoticallySafeQuantumGravity}

\subsection{Classical Action} 
\label{sec:ClassicalAction}

The classical action for gravity is given by the Einstein-Hilbert (EH) action
\begin{align}
    S_\text{EH} [g]=-\frac{1}{16\pi G_\text{N}}\int \!\mathrm d^4x\sqrt{g}\, (R-2\Lambda)+ S_\text{gf} + S_\text{gh}\,,
\label{eq:EHAction}
\end{align}
where $G_\text{N}$ is Newton's constant, $\Lambda$ is the cosmological constant, $R$ is the Ricci scalar, and $g$ is the absolute value of the determinant of the metric, $g = |\text{det}\,g_{\mu\nu}|$. The action is augmented with a gauge-fixing and ghost action, given in \Cref{app:GaugeFixing}. We work in Euclidean signature in the first place, and we will perform a reconstruction of the Lorentzian result later. We use the background field method for the quantisation procedure of \labelcref{eq:EHAction}, where the full metric $g_{\mu\nu}$ is split into a background metric $\bar{g}_{\mu\nu}$ and a fluctuation field $h_{\mu\nu}$ according to
\begin{align}
    g_{\mu\nu} = \delta_{\mu\nu} + \sqrt{G_\text{N}}\ h_{\mu\nu}\,,
\label{eq:MetricSplit}
\end{align} 
where we have chosen the background metric to be the flat Euclidean metric $\delta_{\mu\nu}$. The factor $\sqrt{G_\text{N}}$ in front of the fluctuation field $h_{\mu\nu}$ is introduced for dimensional reasons, since it renders $h_{\mu\nu}$ a field of mass dimension 1. 

In the matter sector, we consider a single massless scalar field $\phi$ minimally coupled to gravity, whose classical action reads
\begin{align}
    S_{\phi}[g,\phi]=\frac{1}{2}\int \!\mathrm d^4x\sqrt{g}\, g^{\mu\nu} \partial_\mu\phi \,\partial_\nu\phi\,.
\label{eq:ScalarAction}
\end{align}
Note that despite its simplicity, the action \labelcref{eq:ScalarAction} already contains an infinite number of interaction terms $h^n \phi^2$ between the scalar field and the graviton, due to the presence of $\sqrt{g}$ and $g^{\mu\nu}$.

\subsection{Scalar Scattering}
\label{sec:ScalarScattering}
 
\begin{figure*}[t]
	\centering
	\includegraphics[width=0.9\textwidth]{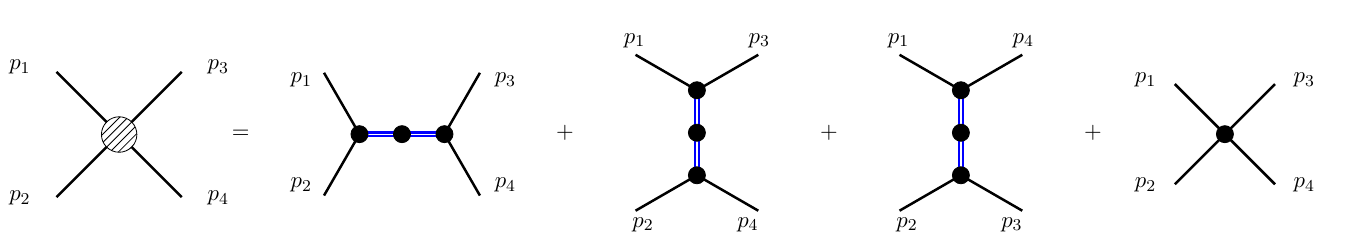}
	\caption{Feynman diagrams for the scattering of two identical scalars into two identical scalars, see \Cref{eq:FullChannelsAmplitude}. All vertices and propagators are full 1PI vertices, see \Cref{sec:FRG}. In this work, we focus on the first three mediated diagrams, corresponding to $s,t,u$-channels.}
	\label{fig:scattering}
\end{figure*}

The scattering of identical scalars is shown in \Cref{fig:scattering} with full 1PI vertices. In general, the amplitude in the symmetric phase is composed of four diagrams, the $s$, $t$, and $u$ channel diagrams as well as the direct contact term,
\begin{align}
	\mathcal{A} = \mathcal{A}_{s} + \mathcal{A}_{t} + \mathcal{A}_{u} + \mathcal{A}_{4}\,,
\label{eq:FullChannelsAmplitude}
\end{align}
where $s,t$ and $u$ are the Mandelstam variables. In this work, we focus on the graviton-mediated diagrams and neglect the direct contact term $\mathcal{A}_{4}$, whose computation is subject to \cite{PCA2026}. Note that a leading order result of $\mathcal{A}_4$ in asymptotic safety has recently been computed in \cite{Knorr:2026vax}.

The tree-level amplitude for $\mathcal{A}_{s}$ is obtained from \Cref{fig:scattering} with the classical scalar-graviton vertex in   \labelcref{eq:ScalarAction} and the classical graviton propagator in \labelcref{eq:EHAction}. The result is given by
\begin{align}
    \mathcal{A}_{s} = \frac{G_\text{N}}{s}\left[ (t^2 - 4tu + u^2) - s^2 \right],
\label{eq:ClassicalAmplitude}
\end{align}
and the $t$ and $u$ channel amplitudes are obtained through crossing symmetry, $\mathcal{A}_t=\mathcal{A}_s(s\leftrightarrow t)$ and $\mathcal{A}_u=\mathcal{A}_s(s\leftrightarrow u)$. Note that we are interested in the high-energy behaviour of the amplitude, and thus we have assumed a vanishing scalar mass.

This result exhibits two fundamental pathologies:
\begin{itemize}
\item Unitarity violation in the UV, as the amplitude grows linearly with the centre-of-mass energy, $\mathcal{A}_{s} \sim G_\text{N}\, s$.
\item A forward-scattering divergence proportional to $G_\text{N} \, s^2/t$ for $s \to \infty$ at fixed $t$.
\end{itemize}
These issues are linked to the negative mass dimension of the Newton constant $[G_\text{N}] = -2$, which is at the heart of the perturbative non-renormalisability of general relativity. In fact, beyond tree level, the UV scaling of \labelcref{eq:ClassicalAmplitude} is worsened by perturbative loop corrections. A consistent UV completion of gravity must resolve both unitarity and renormalisability simultaneously.

We want to study the non-perturbative scattering amplitude in the framework of asymptotically safe quantum gravity. The full non-perturbative amplitude is obtained by replacing the classical vertices with full 1PI vertices. Schematically, it can be written as 
\begin{align}
    \mathcal{A}_{s} = \frac{G_\text{N}}{s}\left[ V_2(s)(t^2 - 4tu + u^2) - V_0(s)s^2 \right].
\label{eq:1PIAmplitude}
\end{align}
The amplitude \labelcref{eq:1PIAmplitude} already contains two remarkable features: First, the full non-perturbative information is contained in two dressing functions $V_{i}$ that relate to the contribution from the spin-$i$ mode of the graviton. Second, these dressing functions depend only on the Mandelstam variable of the respective channel. 

We want to compute the functions $V_2$ and $V_0$ non-perturbatively. The functional renormalisation group (fRG) is well suited for this task, in particular fluctuation approach to quantum gravity \cite{Pawlowski:2023gym, Pawlowski:2020qer}, which focuses on flow equations for 1PI correlation functions of the fluctuation field $h_{\mu\nu}$ and matter fields. With this approach, the functions $V_2$ and $V_0$ are composed of the 1PI graviton-scalar vertex and the graviton propagator, as is evident from the diagram, \Cref{fig:scattering}. The graviton propagator was already computed in Lorentzian signature \cite{Fehre:2021eob,Pawlowski:2025etp,Assant2026}, as well as with a reconstruction from Euclidean precision data \cite{Bonanno:2021squ}. In all cases positive spectral functions without ghost poles have been obtained and the results corroborate the qualitative form of the graviton spectral function. Here we focus on the computation of the scalar-graviton vertex $\Gamma^{\phi\phi h}$, which completes the input for the graviton-mediated amplitudes. 

The dressing functions can be related to form factors of the quantum effective action \cite{Knorr:2022dsx, Knorr:2022lzn, Knorr:2021niv, Draper:2020bop, Draper:2020knh}. Specifically, $V_{2}$ is related to the form factor of the Weyl-tensor-squared, $C_{\mu\nu\rho\sigma} f_C(\Box) C^{\mu\nu\rho\sigma}$ and the form factor of  $R^{\mu\nu}\partial_\mu\phi \partial_\nu\phi$, while  $V_{0}$ is related to the form factor of the Ricci-scalar-squared, $R f_R (\Box) R$, also including the form factors associated with $R^{\mu\nu}\partial_\mu\phi \partial_\nu\phi$ and $R\,\partial_\mu\phi \partial^\mu\phi$.

The full amplitude \labelcref{eq:FullChannelsAmplitude} has been computed in a loop expansion within the effective field theory of general relativity \cite{Donoghue:1993eb, Donoghue:1994dn, Bjerrum-Bohr:2002gqz}, and also eikonal resummations, see \cite{tHooft:1987vrq, Amati:1987wq, Amati:1987uf, Muzinich:1987in, Kabat:1992tb} for older string-inspired works and \cite{Giddings:2010pp, DiVecchia:2021bdo, DiVecchia:2023frv} for modern applications. While the high-energy behaviour of the amplitudes needs to be assessed with non-perturbative methods, these results offer crucial benchmarks in the IR. Here, we focus on the graviton-mediated contributions, $\mathcal{A}_{s} + \mathcal{A}_{t} + \mathcal{A}_{u}$, while the gravitational contributions to $\mathcal{A}_4$ will be discussed in a forthcoming work \cite{PCA2026}. The latter have also been recently addressed in \cite{Knorr:2026vax}.

\section{Euclidean Quantum Gravity}
\label{sec:EuclideanQuantum gravity}

Asymptotically save quantum gravity is best accessed within the fRG, for a comprehensive review see \cite{Dupuis:2020fhh}, and for a recent collection of reviews on asymptotically safe quantum gravity, see \cite{Bambi:2023jiz}.

\subsection{Functional Renormalisation Group}
\label{sec:FRG}

In the fRG, an infrared (IR) regularisation is introduced to the path integral that suppresses quantum modes below a given scale $k$. This is achieved through a modified dispersion $p^2 \rightarrow p^2 + R_k$, where $R_k$ is the IR regulator that suppresses the propagation of momentum degrees of freedom with $p^2 \lesssim k^2$. This IR regularisation goes hand in hand with a scale-dependent effective action, in our case $\Gamma_k[\bar{g}_{\mu\nu},\Phi]$ with the background metric $\bar g_{\mu\nu} = \delta_{\mu\nu}$ and the dynamical fluctuation fields 
\begin{align}
\Phi=(h_{\mu\nu},c_\mu,\bar{c}_\nu,\phi)\,.
\label{eq:Phi}
\end{align}
The full quantum effective action is obtained for a vanishing regulator, that is $k \to 0$: $\Gamma = \Gamma_{k=0}$. The scale dependence of $\Gamma_k$ is governed by the Wetterich equation \cite{Wetterich:1992yh}, see also \cite{Morris:1993qb, Ellwanger:1993mw}, for gravity see \cite{Reuter:1996cp}, 
\begin{align}
{\partial_t}\Gamma_k[\Phi]=\frac{1}{2}\text{Tr}\left[\frac{1}{\Gamma_k^{(2)}+R_k}\partial_t R_k\right]\,.
    \label{eq:WetterichEquation}
\end{align}
Here, $t=\log(k/k_0)$ is the (negative) RG-time, $k_0$ some reference scale, $\Gamma_k^{(2)}$ the second derivative of the effective action with respect to the fields, and the trace sums over all indices as well as (loop) momenta. In a vertex expansion, the flow equation is a coupled set of integral-differential equations of the $n$-point correlation functions $\Gamma_k^{(n)}$, 
\begin{align}
    \Gamma^{(n)}_{k, \Phi_{m_1}\dots\Phi_{m_n}}(p_1,...,p_n) := \frac{\delta^n \Gamma_k[\Phi]}{\delta \Phi_{m_1}(p_1)\dots\delta \Phi_{m_n}(p_n)}\Bigg|_{\Phi=0}\!\!.
    \label{eq:CorrelationFunctions}
\end{align}
For more details in the context of gravity, see \cite{Pawlowski:2020qer,Pawlowski:2023gym}.

\subsection{Amplitude and Vertices}
\label{sec:AmplitudeAndVertices}

\begin{figure}[tbp]
    \includegraphics[width=0.7\linewidth]{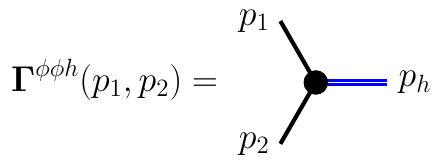}
    \caption{Full 1PI scalar-graviton vertex with $p_h= -p_1 - p_2$. All momenta are considered as incoming.}
    \label{fig:scalar-gravitonVertex}
\end{figure}

We focus on the $s$-channel amplitude ${\cal A}_{s}$ shown in \Cref{fig:scattering}. The amplitudes of the $t$-  and $u$-channels can be obtained through crossing symmetries. As outlined in \Cref{sec:AsymptoticallySafeQuantumGravity}, the amplitude can be readily expressed in terms of the full propagator and full vertices. We denote the 1PI two-scalar--one-graviton vertex by $\Gamma^{\phi\phi h}$, see \Cref{fig:scalar-gravitonVertex}, and the 1PI graviton propagator by $\mathcal{G}_{hh}$. Then the $s$-channel amplitude reads
\begin{align}\nonumber 
    {\cal A}_{s} = &\frac{1}{ \sqrt{Z_\phi(p_1) Z_{\phi}(p_2)} }\, \Gamma^{\phi\phi h}_{\mu\nu}(p_1,p_2) \\[2ex]
    &\times\mathcal{G}_{hh}^{\mu\nu\rho\sigma}(p_h)\, \Gamma^{\phi\phi h}_{\rho\sigma}(p_3,p_4)\, \frac{1}{ \sqrt{Z_\phi(p_3) Z_{\phi}(p_4)} }\,,
\label{eq:AmplitudeAndVertices}
\end{align}
where $p_h = p_1 + p_2 = p_3 + p_4$. We have divided out one wave function $\sqrt{Z_\phi (p_i)}$ for each external scalar leg. Strictly speaking \labelcref{eq:AmplitudeAndVertices} is only an amplitude if evaluated on-shell, where it reduces to the LSZ reduction formula. This requires the Wick rotation to Lorentzian signature.

Note that due to momentum conservation, each vertex is a function of only two independent four-momenta, where the latter is defined with
\begin{align}
\hspace{-.15cm}\Gamma^{(3)}_{\phi\phi h}(p_1,p_2,p_3)=\Gamma^{\phi\phi h}(p_1,p_2) (2 \pi)^4\delta(p_1+p_2 + p_3)\,.
\label{eq:GphiphihKernel}
\end{align}
The structure of the amplitude \labelcref{eq:AmplitudeAndVertices} is simplified by appropriate rescalings of the vertices with the wave function,
\begin{align}
    \bar{\Gamma}^{(n)} (p_1,...,p_n) = \frac{ \Gamma^{(n)}(p_1,...,p_n) }{ \prod_{i=1}^n  \sqrt{Z_{\Phi_{m_i}} (p_i)} } \,,
\label{eq:CorrelationFunctionRgInvariant}
\end{align}
with $Z_{\Phi_{m_i}}$ the wave functions of the $i$-th field, and $\bar{\Gamma}^{(n)}$ is the RG-invariant $n$-point vertex, see \cite{Pawlowski:2020qer, Pawlowski:2023gym}. Accordingly, we define the RG-invariant propagator $\bar{\mathcal{G}}$ as the inverse of the RG-invariant two-point function. 

The amplitude \labelcref{eq:AmplitudeAndVertices} is then conveniently expressed in terms of RG-invariant quantities as
\begin{align}
    \mathcal{A}_{s} = \bar{ \Gamma }^{ \phi\phi h}_{ \mu\nu } (p_1, p_2)\, \bar{ \mathcal{G}}_{hh}^{ \mu\nu\rho\sigma } (p_h)\, \bar{ \Gamma }^{ \phi\phi h}_{ \rho\sigma }(p_3, p_4)\,.
\label{eq:RgInvariantAmplitude}
\end{align}
where all the wave functions $Z_i$ have conveniently cancelled out. This implies that the graviton propagator $\bar{\mathcal{G}}_{hh}$ at $k=0$ is just given by the classical dispersion,
\begin{align} 
    \bar{\mathcal{G}}_{hh}^{ \mu\nu\rho\sigma }  (p) = \frac{1}{p^2}  \mathcal{P}^{\mu\nu\rho\sigma}(p) \,,
\label{eq:graviton-propagator}
\end{align}
where $\mathcal{P}^{\mu\nu\rho\sigma}$ is the tensor structure of the graviton propagator in the harmonic-Landau gauge, see \Cref{app:GravitonPropagator} for details. The wave function $Z_h (p)$, which acts as a propagator dressing, has been fully absorbed into the vertices. 

We parametrise the RG-invariant scalar-graviton vertex with
\begin{align}
    \bar{\Gamma}^{\phi\phi h}_{\mu\nu}(p_1,p_2) = G^{1/2}_{\phi\phi h}(p_1,p_2)\, \mathcal{T}_{\mu\nu}^{\phi\phi h}(p_1,p_2)\,,
\label{eq:VertexParametrization}
\end{align}
where $G^{1/2}_{\phi\phi h}$ is the RG-invariant vertex dressing of the classical tensor structure $\mathcal{T}^{\phi\phi h}$ derived from \labelcref{eq:ScalarAction}. Since this vertex dressing has overlap with the Newton coupling, it is commonly referred to as momentum-dependent avatar of the Newton coupling associated with the vertex. Note that due to the symmetry of the vertex $\bar{\Gamma}^{\phi\phi h}_{\mu\nu}(p_1,p_2) = \bar{\Gamma}^{\phi\phi h}_{\mu\nu}(p_2,p_1)$, also the Newton coupling $G_{\phi\phi h}$ is symmetric under the exchange of momenta, $G_{\phi\phi h}(p_1,p_2)= G_{\phi\phi h}(p_2,p_1)$.

The parametrisation \labelcref{eq:VertexParametrization} contains the first approximation of this work: The scalar-graviton vertex contains three tensor structures, see \labelcref{eq:ScalarTensorStructure}, and, in principle, each tensor structure should be equipped with its own vertex dressing. Here, we only compute the vertex dressing of the classical tensor structure, which has overlap with all propagating modes of the graviton. For the full tensor basis of the $\phi\phi h$ vertex, see \Cref{app:VertexExpansion}.

\begin{figure*}[tbp]
\includegraphics[width=\linewidth]{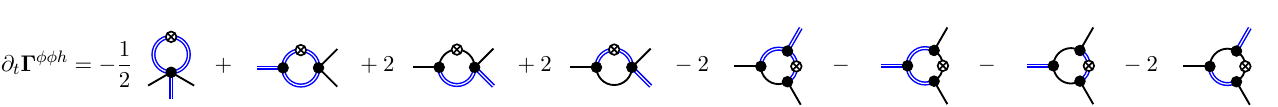}
    \caption{Diagrammatic representation of the flow of the scalar-graviton vertex. The single black lines represent scalar propagators, while the double blue lines represent graviton propagators. The crossed circles stand for regulator insertions. Note the absence of closed matter loops, since we are neglecting scalar self-interactions, see \labelcref{eq:ScalarAction}.}    \label{fig:DiagrammaticFlowEq}
\end{figure*}

In summary, \labelcref{eq:RgInvariantAmplitude} reduces to
\begin{align}\nonumber 
    \mathcal{A}_{s} &= \sqrt{G_{\phi\phi h}(p_1, p_2) \, G_{\phi\phi h}(p_3, p_4)}  \\[2ex]
    &\qquad\times\frac{ \mathcal{T}_{\mu\nu}^{\phi\phi h}(p_1, p_2) \,\mathcal{P}^{\mu\nu\rho\sigma} (p_h)\, \mathcal{T}_{\rho\sigma}^{\phi\phi h}(p_3, p_4) }{p_h^2}\,,
\label{eq:AmplitudeRgInvCancellations}
\end{align}
and we are left with an amplitude where all non-trivial information is stored in the vertex dressing $G_{\phi\phi h}$. The tensor structures in the second line can be evaluated straightforwardly, and they give
\begin{align}
   \mathcal{T}_{\mu\nu}^{\phi\phi h}\, \mathcal{P}^{\mu\nu\rho\sigma} \, \mathcal{T}_{\rho\sigma}^{\phi\phi h} =  t^2 - 4tu + u^2 - s^2 ,
\label{eq:TensorContractionStu}
\end{align}
in terms of the Mandelstam variables. Trivially, this agrees with the tree-level amplitude given in \labelcref{eq:ClassicalAmplitude}. The remaining factors in \labelcref{eq:AmplitudeRgInvCancellations} can all be expressed in terms of the Mandelstam $s$. Naturally, \labelcref{eq:AmplitudeRgInvCancellations} is of the same form as \labelcref{eq:1PIAmplitude}. It almost fully resolves the structure of the amplitude, apart from the approximation $V_2(s) = V_0 (s)$, which is due to our approximation to resolve only one vertex dressing in \labelcref{eq:VertexParametrization}.

From \labelcref{eq:AmplitudeRgInvCancellations} it is clear that the computation of the non-perturbative vertex $\Gamma^{\phi\phi h}$ is tantamount to the computation of the non-perturbative amplitude ${\cal A}_{s}$. In the following, we provide the first genuine computation of the momentum dependence of a matter-gravity vertex $\Gamma^{\phi\phi h}$. For this, we integrate the flow equation for $\partial_t \Gamma^{\phi\phi h}$, whose diagrammatic form is shown in \Cref{fig:DiagrammaticFlowEq}. We denote the projected flow of $\Gamma^{\phi\phi h}$ with 
\begin{align}
    F_k\!\left(p_1,p_2; \{ g_k^{(n)} \} \right) := 2\ \frac{ \partial_t \bar{\Gamma}^{\phi\phi h}_{k,\mu\nu} \mathcal{T}^{\phi\phi h, \mu\nu} }{ \mathcal{N} }\,,
\label{eq:ShorthandOfFlow}
\end{align}
where the explicit presence of $g_k^{(n)}$ accounts for the different avatars of the Newton coupling (vertex dressings), see \Cref{fig:DiagrammaticFlowEq}. Furthermore, the normalisation factor of the projection procedure $\mathcal{N}$ is given by
\begin{align}
    \mathcal{N} := k^{-1}\, \mathcal{T}^{\phi\phi h, \mu\nu } \mathcal{T}^{\phi\phi h}_{\mu\nu}\, ,
\label{eq:NormDefinition}
\end{align}
and the explicit factor of 2 in \labelcref{eq:ShorthandOfFlow} stems from the $\sqrt{g_{\phi\phi h,k}}$-derivative of the left-hand side of the flow equation. The flow function $F_k$ was computed from the diagrams shown in \Cref{fig:DiagrammaticFlowEq} with the help of self-written codes that are based on FORM \cite{Vermaseren:2000nd, Kuipers:2012rf} and are advertised in the asymptotic safety codebase \cite{AScodebase}.

Using the vertex parameterisation in \labelcref{eq:VertexParametrization}, we get a differential equation of the form
\begin{align}\nonumber 
    \partial_t g_{\phi\phi h,k} (p_1, p_2) &=
    \left( 2+ \sum_j \eta_{\Phi_j}(p_j) \right) g_{\phi\phi h,k} (p_1, p_2)  \\[2ex]
    &\hspace{-.5cm}+ g^{1/2}_{\phi\phi h,k} (p_1, p_2) \, F_k\!\left(p_1,p_2; \{ g_k^{(n)} \} \right)\,,
\label{eq:scalar-gravitonFlow}
\end{align}
where $g_{\phi\phi h,k} = G_{\phi\phi h,k} k^2$ is the dimensionless Newton coupling, and $\eta_{\Phi_j}$ are the anomalous dimensions of the fields defined as
\begin{align}
    \eta_{\Phi_j}(p) = -\partial_t \ln Z_{\Phi_j,k}(p)\,.
\label{eq:anomalous dimensions}
\end{align}
Note that \labelcref{eq:scalar-gravitonFlow} is written in terms of dimensionful 4-momenta $p_i$ and one could equivalently switch to dimensionless 4-momenta $\hat p_i = p_i/k$, thus picking up additional $\hat p_i$-derivatives. The details of the derivation of \labelcref{eq:scalar-gravitonFlow} are provided in \Cref{app:FlowEquationDerivation}. Furthermore, in \Cref{app:projection of flow} we discuss the projection schemes.

\subsection{Approximations and Flow} \label{sec:ApproximationsAndConventions}

We now construct approximations to the full flow equation \labelcref{eq:scalar-gravitonFlow}, that facilitate the numerical integration of the flow significantly: 
the most challenging property of \labelcref{eq:scalar-gravitonFlow} is the dependence of the flow on the various momentum-dependent avatars of the Newton coupling, $g_k^{(n)}$. 
Moreover, these couplings depend on both the loop momentum $q$ and the external momenta $p_1, p_2$. 
Consequently,  \labelcref{eq:scalar-gravitonFlow} is an integral-differential equation, coupled to integral-differential equations for the other Newton couplings. 
Since we are not resolving the other avatars of the Newton couplings, we make the approximation to identify all avatars of the Newton coupling with $g_{\phi\phi h,k}$. 
In \cite{Eichhorn:2018akn,Eichhorn:2018ydy}, the difference between different three-point avatars of the Newton coupling was studied. 
It was found that the avatars are approximately equal at the UV fixed point, a property called effective universality. Moreover, in the deep IR (far below the Planck scale) the effective action is governed by perturbation theory, and the couplings are identical up to subleading terms due to the Slavnov-Tayor identities. This motivates our approximation to globally identify all avatars for all cutoff scales.

Within this approximation, \labelcref{eq:scalar-gravitonFlow} is a closed integral-differential equation for the coupling $g_{\phi\phi h}(p,z)$. It still is an integral-differential equation, and we invoke a further approximation in order to factorise the momentum structure. For that purpose we use that the loop momentum $q$ is restricted by the cutoff scale, $q^2\leq k^2$, and the vertices or rather the respective couplings only show a mild momentum-dependence for these momenta. For respective discussions see in particular \cite{Denz:2016qks, Ihssen:2024miv}. A similar reasoning also underlies the BMW-scheme \cite{Blaizot:2005xy}. Consequently, we evaluate the loop momentum in the  couplings at $q=0$, and arrive at 
\begin{align}
    F_k\!\left(p_1,p_2; \{ g_k^{(n)} \} \right) \approx g_{\phi\phi h, k}^{3/2}(p_1, p_2)\, F_k(p_1, p_2; 1).
\label{eq:FlowAvatarIdentification}
\end{align}
%
%
\begin{figure*}[tbp]
	\centering
    \includegraphics[width=.45\linewidth]{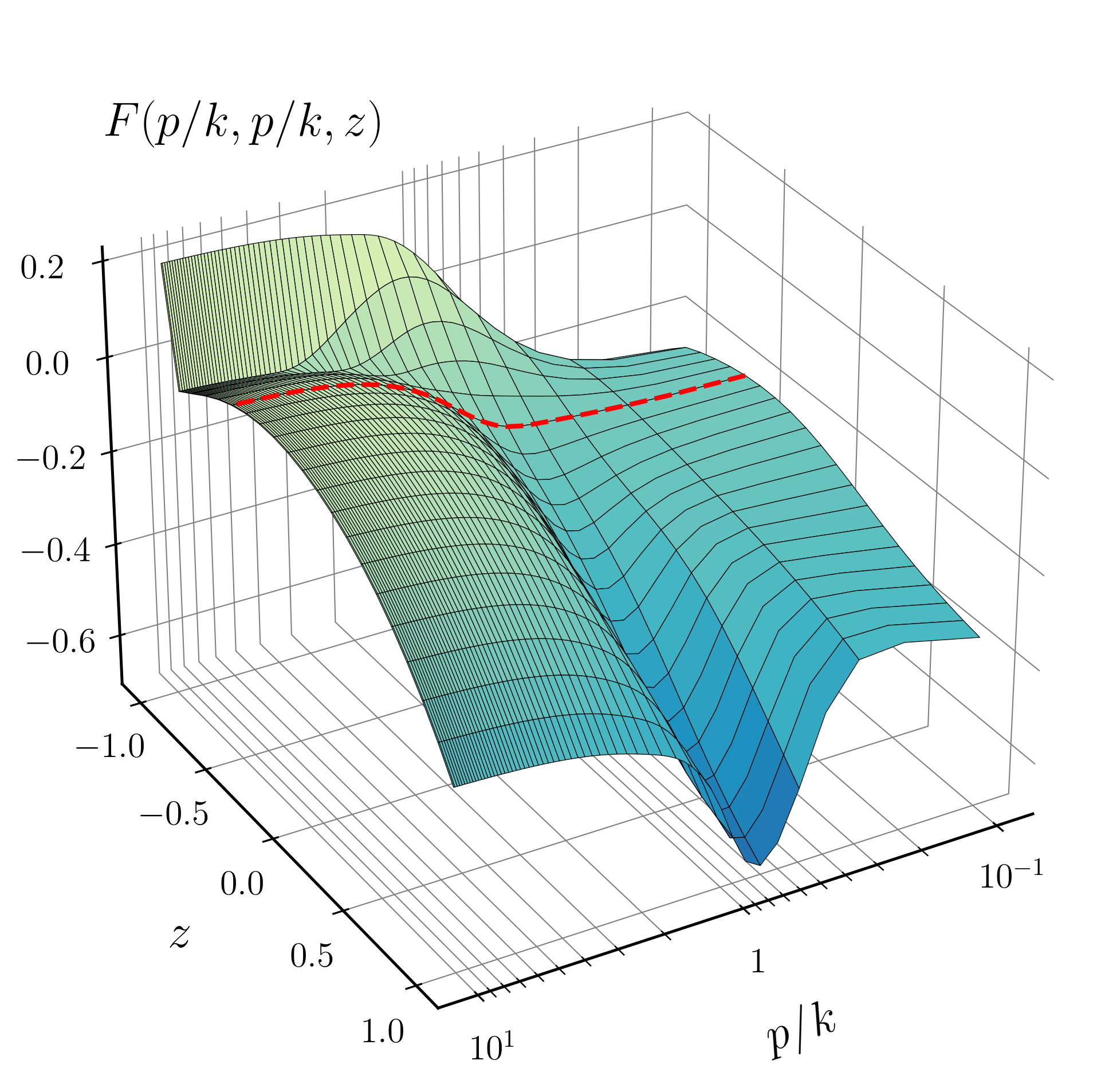}
    \hspace{1.2cm}
    \includegraphics[width=.45\linewidth]{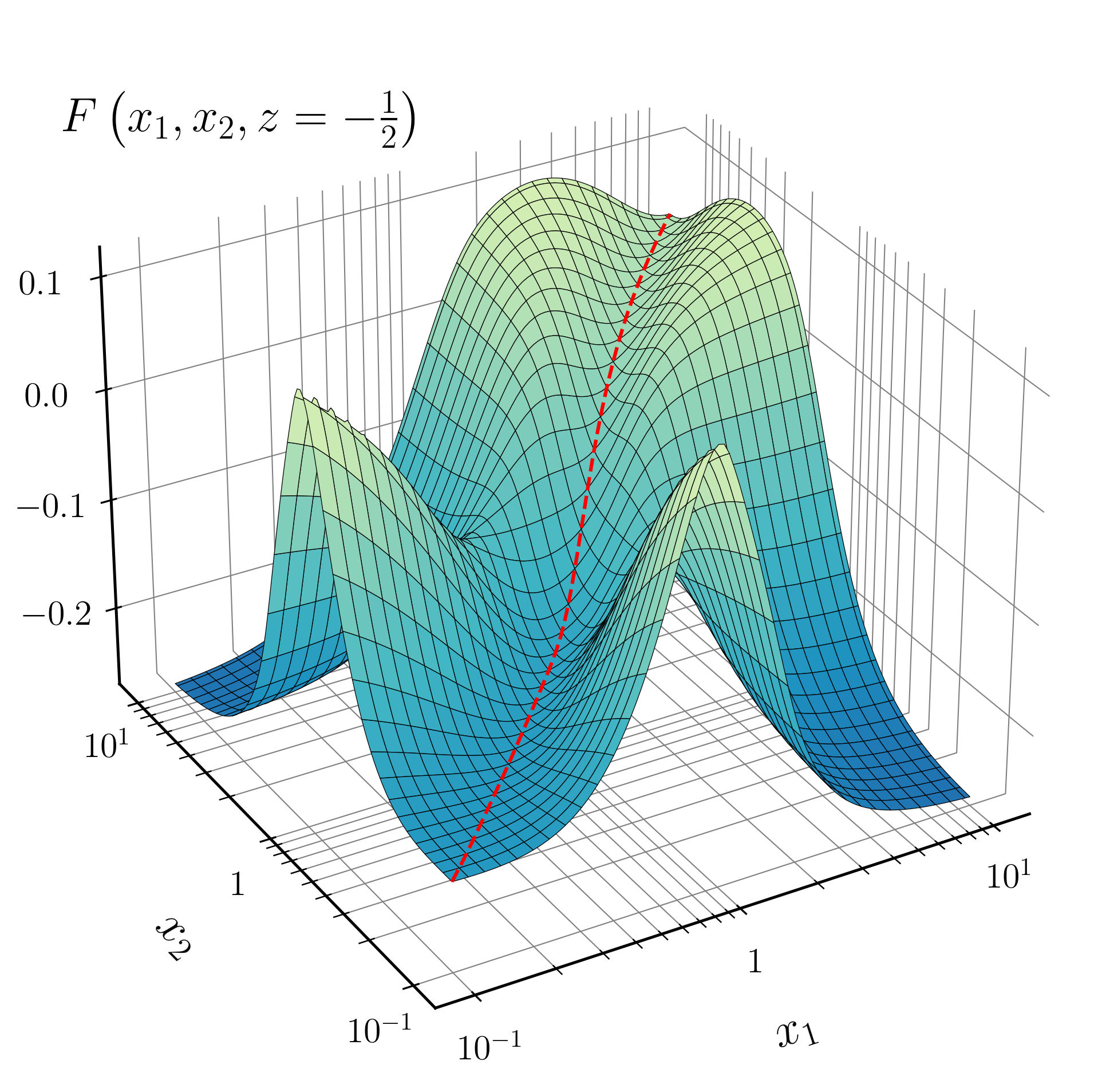}
    \caption{Flow function $F(x_1,x_2,z)$, \labelcref{eq:FlowFunctionP1P2andZ}, of the scalar-graviton vertex defined in \labelcref{eq:ShorthandOfFlow} for generic radial momenta $x_1=\|p_1\|/k\, , x_2=\|p_2\|/k\,$ and $z$, see \labelcref{eq:Anglez}. $F(x_1,x_2,z)$ is the diagrammatic (dynamical) part of the flow of the Newton coupling  \labelcref{eq:FlowEquationFinalFinal}. Left: projection onto the $p/k=x_1=x_2$ plane. It is shown as a function of the dimensionless momentum $p/k$ and the angle $z$. Right: projection onto the momentum symmetric plane $z=-1/2$ for independent $x_1,x_2$. In both figures, the red dashed lines are the slices $x_1=x_2$ with $z=-1/2$.
    }
    \label{fig:FlowResolution}
\end{figure*}
%
Using \labelcref{eq:FlowAvatarIdentification}, the flow \labelcref{eq:scalar-gravitonFlow} reduces to an ordinary differential equation for the coupling $g_{\phi\phi h,k}(p_1,p_2)$, 
\begin{align}\nonumber 
     \partial_t g_{\phi\phi h,k}(p_1,p_2) = &\Bigl[ 2+\eta_h(p_1+p_2)\\[1ex]\nonumber 
     & \hspace{-.5cm}+ \eta_\phi(p_1)+\eta_\phi(p_2)\Bigr]\,g_{\phi\phi h,k}(p_1,p_2) \\[1ex]
     & \hspace{-.5cm}+ g_{\phi\phi h,k }^{2}(p_1,p_2) F_k(p_1,p_2;1)\,, 
 \label{eq:FlowEquation-gFull}
\end{align}
where the last line is the diagrammatic contribution in \Cref{fig:DiagrammaticFlowEq}. 

In view of the reconstruction task as well as for further simplification of the computational task, we use further approximations as well as using existing results as input:
\begin{enumerate}
\item In our setup with a  minimally coupled matter-gravity in the harmonic-Landau gauge, the scalar anomalous dimension vanishes exactly for all momenta, $\eta_\phi (p) =0$, see \cite{Meibohm:2015twa}. This entails $Z_\phi=1$ for all $k$.
\item We use a uniform  graviton wave function  $Z^{(a)}_h$, which we identify with the transverse-traceless one, that is $Z^{(tt)}_{h} \equiv Z_h$. 
    \item In the flow of the graviton two-point function, we consider only the gravitational contributions, neglecting the influence of matter, i.e., we set $N_s=0$ and work in the quenched approximation. This is a good approximation for small numbers of $N_s$ \cite{Meibohm:2015twa, Eichhorn:2018akn}. In the flow of the fluctuation anomalous dimension $\eta_h$, we keep the dependence  on the Newton coupling avatar from the three-graviton vertex $g_3$, for which we use a simple trajectory
\begin{align}
        g_3(k) = \frac{g_3^* k^2}{k^2 + g_3^* M_{\text{pl}}}\,,
    \label{eq:g3Trajectory}
    \end{align}
    with $g_3^* = 1$.
\item We neglect the contribution of the anomalous dimensions $\eta(p)$ in $\partial_t R_k$ in the loops. In the current system they have been found to be subleading, see \cite{Meibohm:2015twa}, and we use  
\begin{align}
\hspace{.7cm}        \eta_h(q)&=0\,,
        &
        \eta_c(q)&=0\,,
        & 
        \eta_\phi(q)&=0\,.
        \label{eq:EtaTo0}
\end{align}
\item The flow also depends on two momentum-independent avatars $\lambda_{2,3}$ of the cosmological constant, extracted from the graviton two-point and three-point functions: $\lambda_2 = -\mu/2$ and $\lambda_3$. If the system is evaluated on the equations of motion (on-shell), it exhibits a massless graviton. This can be emulated by $\mu=0$. Moreover, $\lambda_3$, computed on a flat background, has been found to be very small, $\lambda_3\approx 0$, see \cite{Denz:2016qks}. In combination, this suggests using  
\begin{align}
        \mu &=0 \,,& \lambda_3&=0\,.
\label{eq:ParametersTo0}
\end{align}
\end{enumerate}
The Newton coupling $g_{\phi\phi h,k}$ is a function of two independent four-momenta, since one momentum can be eliminated through momentum conservation, $p_h+p_1+p_2=0$. Together with  $O(4)$-symmetry, this reduces the number of independent momentum components from 12 ($4\times3$ components) to 3.  We choose these 3 independent variables to be the radial momenta $\|p_1\|, \|p_2\|$, as well as the $O(4)$ invariant angle
\begin{align}
z = \cos\varphi = \frac{ p_{1} \cdot \,p_{2} } { \|p_1\|\,\|p_2\| }\,.
\label{eq:Anglez}
\end{align}
For the full momentum conventions, see \Cref{app:MomentumConventions}.
In the present approximation, and in particular with \labelcref{eq:ParametersTo0}, the dimensionless flow function can only depend on the ratio $\|p_i\|/k$ and we find 
\begin{align}
     F(x_1,x_2,z)&=F_k(p_1,p_2; 1)\,, 
     &
     x_i &= \frac{\| p_i\|}{k}\,, 
\label{eq:FlowFunctionP1P2andZ}
\end{align}
with the $k$-independent dimensionless function $F(x_1,x_2,z)$. The diagrammatic form of $F(x_1,x_2,z)$ is provided with \Cref{fig:DiagrammaticFlowEq} with $g_k^{(n)}=1$ for all avatars of the Newton coupling. In \Cref{fig:FlowResolution} we display $F(x_1,x_2,z)$ for a generic angle $z$ and for $\|p_1\|=\|p_2\|$  (left), as well as for the angle $z=-1/2$ for independent magnitudes $x_1,x_2$ (right). 

In the left panel of \Cref{fig:FlowResolution}, we see that for $p/k \to 0$, the flow function $F(0,z)$ is negative for all $z$; this ensures that there is a positive UV fixed point $g^*_{\phi\phi h,k}(p\to0,z)$ in the momentum-independent system for all angles $z$, one of the requirements for asymptotic safety. The flow function $F(p/k,z)$ becomes positive at some finite $p/k>0$. Note that the Euclidean angle $z$ is bounded to the interval $z\in[-1,1]$ and the limits $z=\pm1$ correspond to collinear momenta, while $z=-1/2$ corresponds to the momentum symmetric configuration,
\begin{align}
    \|p_1\|&=\|p_2\|=\|p_h\|\,,
    & 
    p_i\cdot p_j &= \frac{1}{2}(3 \delta_{ij} -1 ) p^2\,.
\label{eq:MomSymm}
\end{align}
We observe that the flow has local maximum around the momentum-symmetric configuration, and the flow function is strongly increasing in the collinear limit $z\to -1$.

In the asymptotic limit of $F(p/k,z)$ for $p/k\to\infty$, the flow function approaches a constant, which implies that we need to be careful about the momentum locality of the flow \cite{Christiansen:2015rva}. For momentum-local flows of correlation functions, the integration of a shell of momentum $p\approx k$ only changes the correlation functions at momenta $p^2\lesssim k^2$. For flow functions such as that of the scalar-graviton interaction \Cref{fig:FlowResolution}, a momentum-independent vertex dressing would imply a non-local flow, and therefore taking into account the momentum-dependence of the vertex dressing is of paramount importance to ensure locality, as demonstrated in \Cref{app:localityFlow}. Moreover, in \Cref{app:localityFlow}, we show how a different approximation can spoil the finiteness of correlation functions.

\begin{figure}[tbp]
    \includegraphics[width=\linewidth]{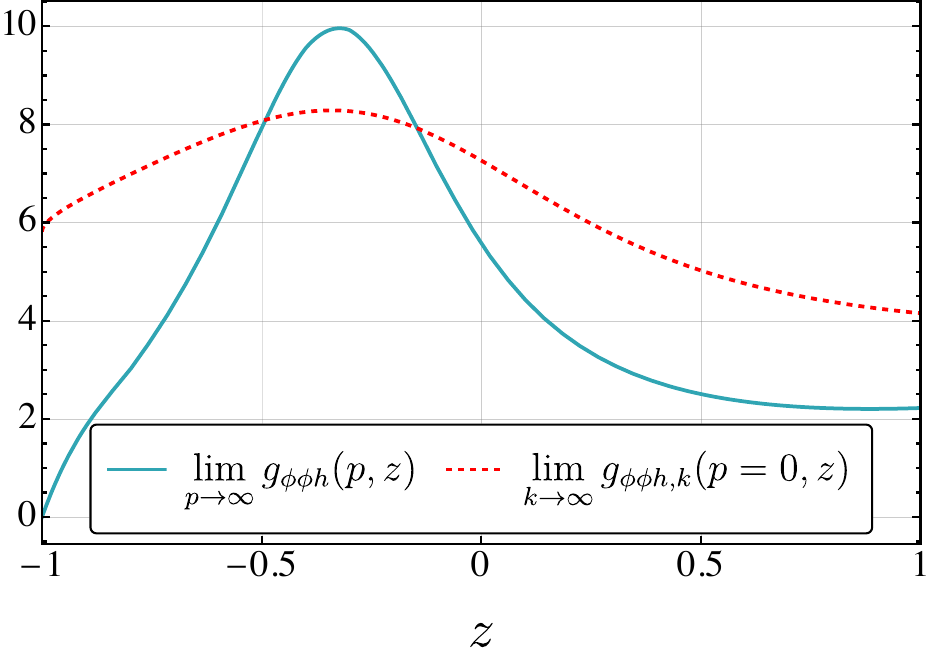}
    \caption{Comparison between the physical momentum fixed point ($k=0$, $p\to \infty$; light blue) and the RG fixed point at vanishing momentum ($k\to \infty$, $p=0$; dashed, red) as a function of the angle $z$. The physical fixed point is obtained from the $p\to\infty$ limit of \labelcref{eq:DimensionlessCouplingP-dependent}, while the momentum-independent one is obtained by solving \labelcref{eq:FlowEquationFinalFinal} at $p=0$.}
    \label{fig:z-dep-fixed-point}
\end{figure}

In the following, we identify the magnitudes of the two scalar momenta, i.e.~$\|p_1\|=\|p_2\|=p$, since this is the relevant momentum configuration for the amplitude. This leaves us with a single momentum variable $p$ and the angle $z$. In summary, the scalar momenta $p_1, p_2$ and the graviton momentum $p_h$ are related to the chosen independent variables $p,z$ as
\begin{align}
    \|p_i\|=p\,,
    \quad  
    \|p_h^2\|=2p^2(1+z)\,,
    \quad 
    z\in[-1,1]\,,
\label{eq:MomentumConventions}
\end{align}
with $i=1,2$. Then,  \labelcref{eq:FlowFunctionP1P2andZ} reduces to a function of two variables,
\begin{align}
    F(p/k, z) := F(p/k,p/k,z)\,.
\label{eq:FlowFunctionPandZ}
\end{align}
This concludes the discussion of the setup and the approximations. 
Using 1.-5.\ and \labelcref{eq:FlowFunctionPandZ} in \labelcref{eq:FlowEquation-gFull}, leads us to
\begin{align}\nonumber 
     \partial_t g_{\phi\phi h,k}(p,z) = &\left[ 2+ \eta_h(2(1+z)p^2) \right]g_{\phi\phi h,k}(p,z) \\[1ex]
     & + g_{\phi\phi h,k }^{2}(p,z) F(p/k,z)\,.
 \label{eq:FlowEquationFinalFinal}
\end{align}
%

\begin{figure*}[htbp]
    \includegraphics[width=\linewidth]{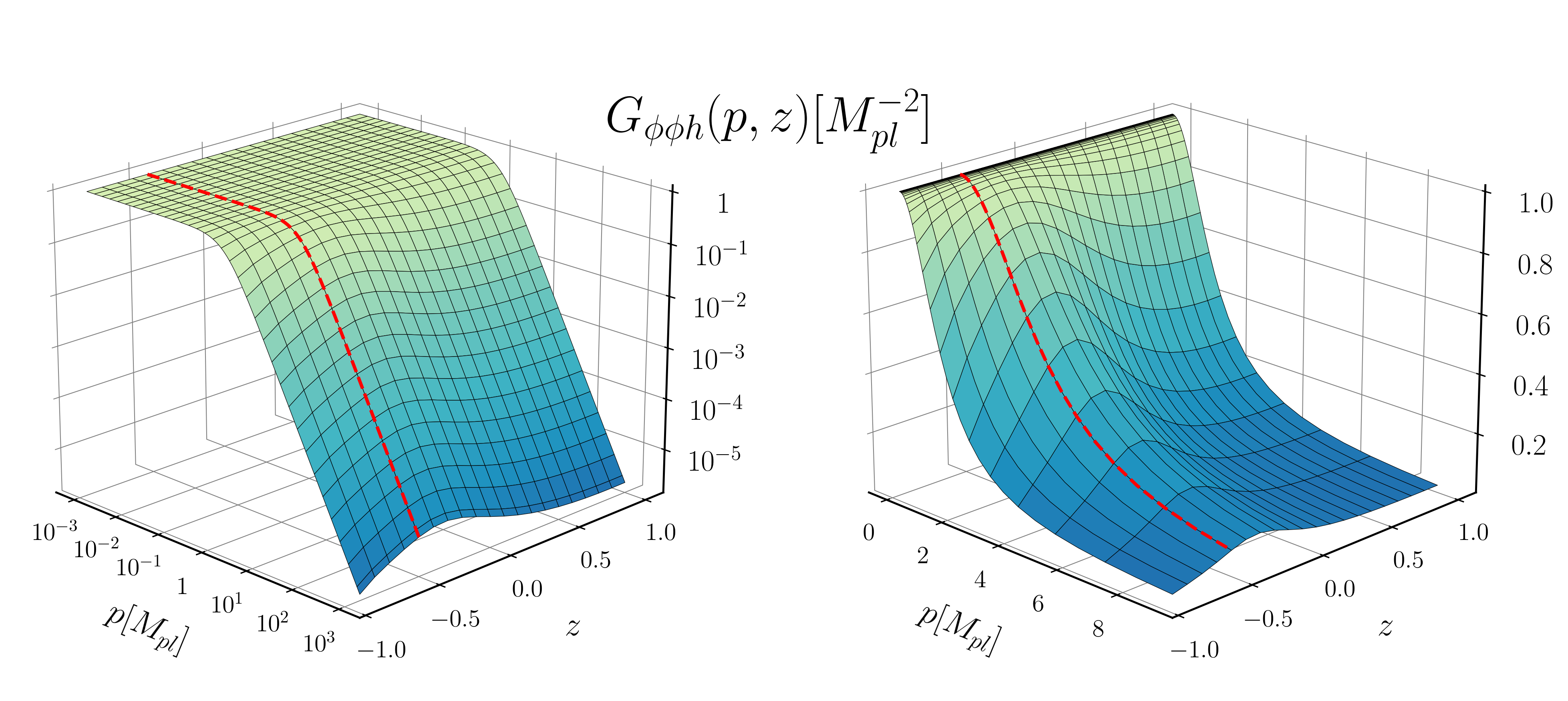}
    \caption{The Newton coupling associated to the scalar-graviton vertex, $G_{\phi\phi h}(p,z)$, as a function of the momentum $p$ (in Planck units) and the angle $z$. It is shown both in logarithmic (left) and linear scale (right). The dashed red line highlights the symmetric momentum configuration. The coupling is asymptotically safe in the UV, scaling as $p^{-2}$, for every angular configuration. Below the Planck scale, $p\ll M_\text{pl}$, we recover classical GR, where $G_{\phi\phi h}=G_\text{N}=M_\text{pl}^{-2}$.}
    \label{fig:FullAngularCoupling}
\end{figure*}

\section{Euclidean Scattering Vertex} \label{sec:ResultsForG_hphiphi}

\subsection{Physical Newton Coupling} \label{sec:NewtonCoupling}

The flow equation for the dimensionless Newton coupling given in \labelcref{eq:FlowEquationFinalFinal} has a $z$-dependent UV fixed point function, which is shown in \Cref{fig:z-dep-fixed-point}. Since we have written \labelcref{eq:FlowEquationFinalFinal} in terms of dimensionful momenta, the limit $k\to \infty$ automatically selects $p=0$. Alternatively, we could write \labelcref{eq:FlowEquationFinalFinal} in terms of dimensionless momenta $\hat p = p/k$. Then the equation would contain an additional $\hat p$-derivative and we would obtain an entire $z$- and $\hat p$-dependent fixed point function.

The critical exponent corresponding to the fixed-point function is $z$-independent and has the value
\begin{align}
    \theta = 3.08\,.
\label{eq:CriticalExponent}
\end{align}
This implies that the coupling is relevant, and we can choose the IR value of the dimensionful Newton coupling to be consistent with general relativity, see below in \labelcref{eq:ClassicalNewtonCoupling}.

From the UV fixed point function, the flow equation \labelcref{eq:FlowEquationFinalFinal} is integrated to $k=0$ on a $(p,z)$-grid and gives the physical dimensionless coupling,
\begin{align}
    g_{\phi\phi h}(p,z) = \lim_{k\to0}\, \frac{p^2}{k^2}\, g_{\phi\phi h,k}(p,z)\,.
\label{eq:DimensionlessCouplingP-dependent}
\end{align}
The quantity entering the scattering amplitude \labelcref{eq:AmplitudeRgInvCancellations} is the dimensionful Newton coupling $G_{\phi\phi h}(p,z)$, which is obtained from \labelcref{eq:DimensionlessCouplingP-dependent} as
\begin{align}
    G_{\phi\phi h}(p,z) = \frac{1}{p^2}\, g_{\phi\phi h}(p,z)\,.
\label{eq:physical coupling}
\end{align}
The result is displayed in \Cref{fig:FullAngularCoupling}, where the physical Euclidean momentum is now measured in units of the Planck scale since the RG-scale $k$ has been integrated out. The logarithmic plot (left) shows the coupling over a wide range of momenta. In particular, we highlight two distinct regimes that corroborate the asymptotic safety scenario:
\begin{enumerate}
\item In the IR ($p\ll M_{\text{pl}}$), the coupling is constant and for any angle $z$ we find 
\begin{align}
        \lim_{p\to0} G_{\phi\phi h}(p,z) = G_\text{N} = \frac1{M_\text{pl}^2}\,,
\label{eq:ClassicalNewtonCoupling}
\end{align}
with $G_\text{N}$ the classical Newton constant. This is the classical regime, where the coupling is scale independent and matches the value of the Newton constant (in Planck units). In this regime, GR dominates and quantum corrections are suppressed.
\item For momenta around the Planck scale and beyond ($p \gtrsim O(M_{\text{pl}})$), gravity becomes strongly coupled. In an EFT setup, this marks the onset of non-renormalisable contributions. In asymptotic safety, it is the onset of quantum scale invariance, which screens gravitational interactions. In other words, the dimensionless coupling hits a fixed point, and the dimensionful one starts to scale as $G_{\phi\phi h}\sim 1/p^2$. This is the hallmark of asymptotic safety, visible in \Cref{fig:FullAngularCoupling}.  
\end{enumerate}
These considerations are independent of the angular configuration $z$. In fact, \Cref{fig:FullAngularCoupling} supports the symmetric point approximation as a reliable one, since the effect of the angular variable is only quantitative and not qualitative. The influence of $z$ is visible on the right-hand side of \Cref{fig:FullAngularCoupling}, where a linear plot is displayed. In practice, while not substantially altering the IR/UV regimes, the angle $z$ influences the crossover region, i.e.~the transition from the classical to the quantum regime. Most importantly, \Cref{fig:FullAngularCoupling} shows that an asymptotically safe Newton coupling exists for all momentum configurations, i.e.~for any given value of $p$ and $z$.

We show the dependence on the angle $z$ in more detail in \Cref{fig:z-dep-fixed-point}, where we compare the physical momentum fixed point, $g_{\phi\phi h}(p\to\infty,z)$, with the RG fixed point at vanishing momentum, $g_{\phi\phi h,k\to\infty}(p=0,z)$. The latter is the one that is usually computed in fixed-point studies and is computationally easier to obtain. Both fixed points show a similar dependence with respect to the angle $z$; in particular, the maximal fixed-point value is located at roughly the same angle. The physical momentum fixed point shows a stronger dependence on the angle than the RG fixed point. Nonetheless, we find that the qualitative agreement between the fixed points is remarkable. Note that the momentum-dependent fixed point in \Cref{fig:z-dep-fixed-point} does not vanish for $z\to-1$. It approaches the value $g_{\phi\phi h}(p\to\infty,z=-1)=0.24$. The small fixed-point value in this limit is related to the positive value of the flow function in this limit, see \Cref{fig:FlowResolution}.

Lastly, we comment on the large fixed-point values found in \Cref{fig:z-dep-fixed-point} compared to other computations in the literature. This is based on two reasons. First, we use the approximation $\mu=0$, which leads to larger values of $g^*$. Second, we use the Newton coupling from the graviton-scalar vertex. In \cite{Eichhorn:2018akn}, it was shown that the fixed-point coupling from the graviton-scalar vertex and the three-graviton vertex are approximately equal in the regime $0.2<p/k<1$, which is a property called effective universality \cite{Eichhorn:2018akn, Eichhorn:2018ydy}. For smaller values of $p/k$, the avatars do not agree, and the Newton coupling from the graviton-scalar vertex is significantly larger. The combination of these two effects leads to the large fixed-point value of \Cref{fig:z-dep-fixed-point}. Finally, we point out that we also use the classical tensor structure to project on the flow, while the projection operator in \cite{Eichhorn:2018akn} also includes a transverse-traceless projection operator. The classical tensor projection leads to comparatively larger fixed point values, see also \Cref{app:projection of flow}.

\Cref{fig:FullAngularCoupling} represents one of the main results of this work. It shows how asymptotic safety is realised in the physical momentum dependence of the Newton coupling in a matter-gravity system, or, more precisely, in the momentum dependence of the vertex dressing. This was achieved by solving the momentum running of the Newton coupling $G_{\phi\phi h}$. Previous computations showed how AS can be realised at the level of the gravitational vertices $G_{h^3}, G_{h^4}$, mostly in the momentum symmetric configuration \cite{Bonanno:2021squ, Pawlowski:2023dda}. We have expanded on this with an extra momentum variable $z$. Its influence does not alter the classical regime nor the scale-invariant one, but it modulates the crossover region. Furthermore, we now have access to all momentum configurations beyond the symmetric point, see \Cref{fig:FlowResolution}, and here we solve the flow equation for the dependence on $p$ and $z$, which is the relevant momentum configuration for the scattering amplitude.

\section{Wick rotation via reconstruction} 
\label{sec:Scalar-gravitonVertexAmplitude}

One of the main challenges of non-perturbative QFTs is the formulation of the theory at timelike momenta or, equivalently, in Lorentzian signature. In fact, most fRG computations to date, and consequently most AS results, are carried out in backgrounds with Euclidean signature. Only in recent years, functional methods on backgrounds with Lorentzian signature have become amenable, see \cite{Fehre:2021eob, Braun:2022mgx, Pawlowski:2025etp, Assant2026} for the spectral fRG and its application to quantum gravity. Moreover, we refer to \cite{DAngelo:2022vsh, Banerjee:2022xvi, DAngelo:2023tis, DAngelo:2023wje, Banerjee:2024tap, Thiemann:2024vjx, Ferrero:2024rvi, DAngelo:2025yoy} for other locally covariant approaches and \cite{Manrique:2011jc, Rechenberger:2012dt, Biemans:2016rvp, Biemans:2017zca, Knorr:2018fdu, Eichhorn:2019ybe, Knorr:2022mvn,  Saueressig:2023tfy, Korver:2024sam, Saueressig:2025ypi} for works based on the ADM decomposition.

For the case of gravity, the Wick rotation is possibly worsened by the dynamical nature of the metric itself, as shown in \cite{Baldazzi:2018mtl, Baldazzi:2019kim}. There are multiple approaches to circumvent this issue, we mostly follow \cite{Bonanno:2021squ}, where a reconstruction of the Lorentzian correlation functions is performed from the Euclidean ones. This is done by means of spectral representations, as well as suitable ansatze for the spectral functions. We refer to \cite{Bonanno:2021squ} for a more detailed discussion on this point.

\subsection{Euclidean Amplitude} \label{sec:ScatteringAmplitude}
We now proceed to derive a scattering amplitude for the process $\phi\phi\to h \to\phi\phi$, as depicted in \Cref{fig:scattering}. In what follows, we restrict the analysis to the $s$-channel, where the incoming four-momenta are $p_1$ and $p_2$, and the outgoing four-momenta are $p_3$ and $p_4$. The graviton momentum is denoted by $p_h$. The $t$- and $u$-channels can be obtained by crossing symmetry. A brief discussion of other channels will be given in \Cref{sec:BeyondS-channel}.

Let us denote $(p_{ \text{in} },z_{ \text{in} })$ the incoming independent components and $(p_{ \text{out} },z_{ \text{out} })$ the outgoing ones, see also \labelcref{eq:MomentumConventions}. Recall that from \labelcref{eq:AmplitudeRgInvCancellations} the amplitude $\mathcal{A}_{s}$ can be expressed in terms of the couplings as
\begin{align}\nonumber 
    \mathcal{A}_{s} =&  \sqrt{G_{\phi\phi h} ( p_{ \text{in} }, z_{ \text{in} } ) \, G_{\phi\phi h} (p_{ \text{out} },z_{ \text{out} })}\\[1ex]
& \hspace{3cm} \times \frac{t^2 - 4tu + u^2 - s^2}{s}\,.
    \label{eq:AmplitudeContraction}
\end{align}
The $s^2$ term stems from the scalar part of $\mathcal{G}_{hh}$, while the remaining addends stem from the transverse-traceless part. We are then left with two distinct pieces, a classical part that is known fully analytic, $(t^2 - 4tu + u^2 - s^2)/s$, and a quantum part that we have computed numerically, $\sqrt{G_{\phi\phi h} (p_{ \text{in} }, z_{ \text{in} })\, G_{\phi\phi h} (p_{ \text{out} },z_{ \text{out} }) }$. The quantum piece encodes the momentum running of the vertex and contains the information of the UV fixed point.

\begin{figure*}[tbp]
    \includegraphics[width= 0.67\columnwidth]{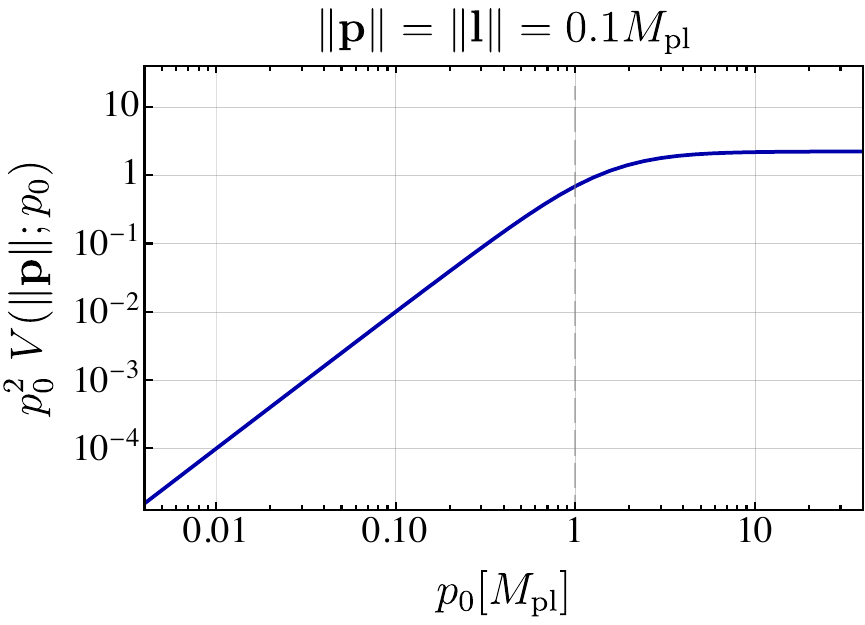}\hspace{.175cm}
    \includegraphics[width= 0.67\columnwidth]{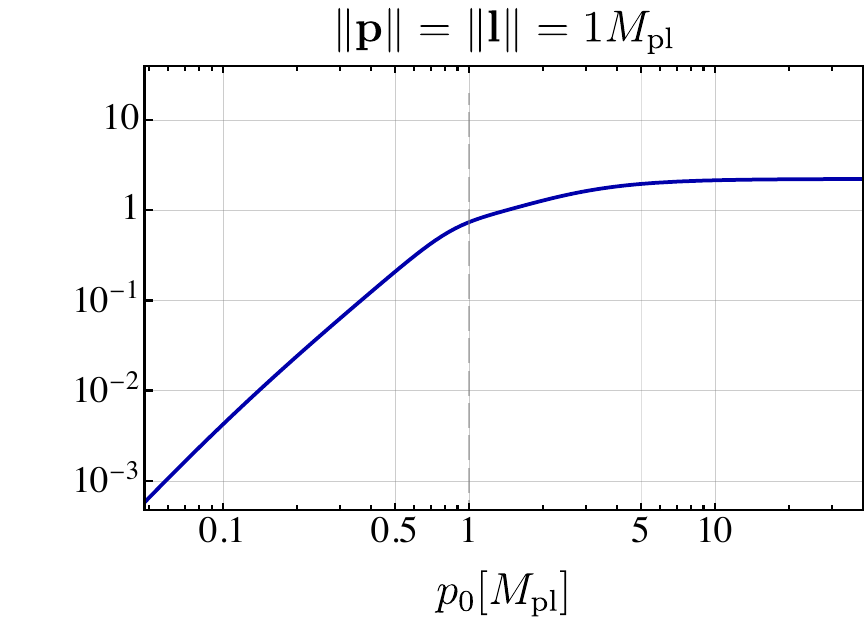}\hspace{.175cm}
    \includegraphics[width= 0.67\columnwidth]{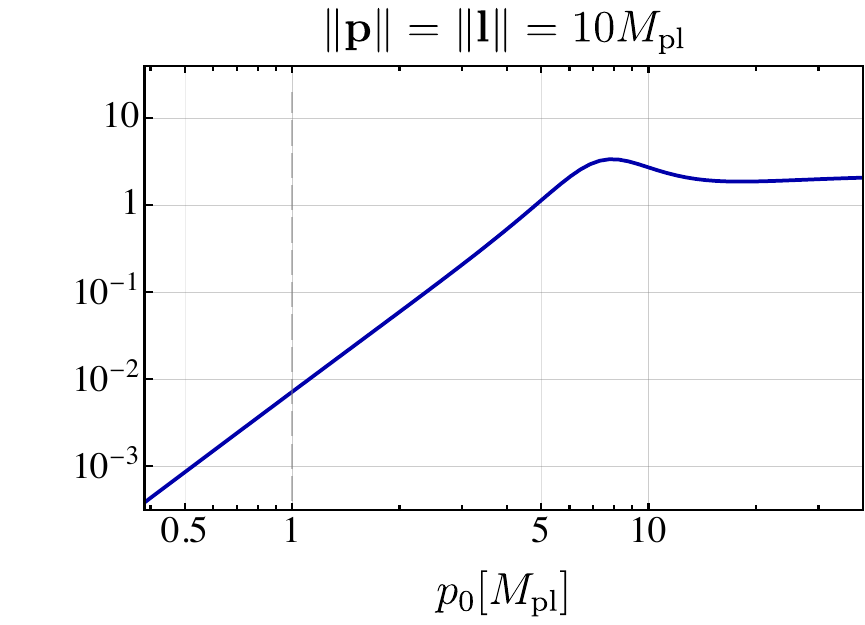}
    \caption{Rescaled vertex $p_0^2\,V(\|\mathbf p\|; p_0)$ as defined in \labelcref{eq:VertexReconstructing} as a function of the frequency $p_0$, for increasing values of $\|\mathbf p\|=\|\mathbf l\|$. The angles are set to $\phi_{\text{in}}=\phi_{\text{out}}=\pi$, i.e.~the centre of mass frame.}
    \label{fig:VertexExternalP}
\end{figure*}

\subsection{Momentum Variables}
\label{sec:MomentumVariables}

The analytical continuation of \labelcref{eq:AmplitudeContraction} is most conveniently done at the level of the temporal component of the four-momentum: given $p_E^2=p_0^2 + \mathbf p^2$, we want to analytically continue $p_0\to -i p_0$, such that $p_L^2=-p_0^2 + \mathbf p^2$ holds. This mimics the signature change of the metric. It was successfully done in \cite{Bonanno:2021squ} for the two-point function where a single momentum is involved. In our case, we have two independent four-momenta, reduced to two independent momentum components $(p,z)$ for each vertex. Moreover, the Euclidean amplitude $\mathcal{A}_s$ \labelcref{eq:AmplitudeContraction} is necessarily off-shell. We have also mentioned that while the parameterisation via $(p,z)$ facilitates the integration of the flow enormously, it hinders a clear analytical continuation. In fact, the $O(4)$-angle $z$ was defined as
\begin{align}
    z = \frac{p_1 \cdot p_2}{ \|p_1\| \|p_2\| }.
\label{eq:z-Definition}
\end{align}
Thus for any massless limit $\|p_1\|,\|p_2\|\to0$, $z$ is ill-defined. In order to make the frequency dependence apparent, we can separate the timelike and spacelike components, i.e.
\begin{align}
    p_1 &= (p_0\, , \mathbf p)\,,& p_3 &= (l_0\, , \mathbf l )\, ,
\label{eq:4MomentaAndFrequency} 
\end{align}
with similar expressions for $p_2$ and $p_4$ as well.
Recall that the scalar four-momenta were taken with the same intensity, i.e.~$\|p_1\|^2=\|p_2\|^2=p_{ \text{in} }^2$ and define two spatial angles $\phi_{ \text{in} }\,, \phi_{ \text{out} }$ such that
\begin{align}
    p_1 \cdot p_2 &= p_0^2 + \mathbf{\|p\|}^2 \cos(\phi_{\text{in}})\,, \notag\\
    p_3 \cdot p_4 &= l_0^2 + \mathbf{\|l\|}^2 \cos(\phi_{\text{out}}).
\label{eq:p1-p2-ScalarProduct}
\end{align}
It then holds that
\begin{align}
    p_{\text{in}}^2  &= p_0^2 + \|\mathbf p\|^2\,,
    &
    z_{\text{in}} &= 1 - \frac{\mathbf{\|p\|}^2}{ \mathbf{\|p\|}^2 + p_0^2} \left(1 -  \cos \phi _{\text{in}}  \right),\notag \\[1ex]
    p_\text{out}^2 &= l_0^2 + \mathbf{\|l\|}^2\,,
    &
    z_{\text{out}} &= 1 - \frac{\mathbf{\|l\|}^2}{\mathbf{\|l\|}^2 + l_0^2} \left(1 -  \cos \phi _{\text{out}} \right) ,
\label{eq:NewMomentumVariables}
\end{align}
which are derived from the definitions of $p$ and $z$, see \labelcref{eq:MomentumConventions,eq:z-Definition}.

The Mandelstam variables $s,t,u$ are related to the new set:
\begin{align}\nonumber 
    s &= 2p^2_{\text{in}} (1 + z_{\text{in}}) = 2p^2_{\text{out}} (1 + z_{\text{out}})\,,\\[1ex]\nonumber 
    t &= p_{\text{in}}^2 + p_{\text{out}}^2 + 2(p_0 l_0 + \mathbf{\|p\|} \mathbf{\|l\|} \cos\theta_{\text{sc}} )\,,
    \\[1ex]
    u &= 2p_{\text{in}}^2 + 2p_{\text{out}}^2 -s - t \,,
\label{eq:MandelstamAndFrequencies}
\end{align}
where $\theta_{\text{sc}}$ is the scattering angle $\mathbf p \cdot \mathbf l=\|\mathbf p\| \|\mathbf l\|\cos\theta_{\text{sc}}$.

In total, we end up with seven variables, namely the Euclidean frequencies $p_0$ and $l_0$, the three-momenta $\mathbf p, \mathbf l$ and the angles $\phi_{\text{in}},\phi_{\text{out}},\theta_{\text{sc}}$. In particular, momentum conservation $(p_1+p_2)_\mu =(p_3 + p_4)_\mu$ relates these as
\begin{align}
     p_0^2 &=  l_0^2 \,,&
     \mathbf{\|p\|}^2 \left( 1 + \cos\phi_{\text{in}} \right) &= \mathbf{\|l\|}^2 ( 1 + \cos\phi_{\text{out}} ) \,,
\label{eq:l0-and-p0}
\end{align}
which we use to eliminate $l_0$ and $\mathbf{\|l\|}$. Moreover, only $\theta_{\text{sc}}$ is a physical angle, while $\phi_{\text{in}}$ and $\phi_{\text{out}}$ parametrise different reference frames. From now on, we work in the Centre of Mass (CoM) frame, in which the angles read
\begin{align}
   \phi_{\text{in}} &= \phi_{\text{out}} = \pi,
\label{eq:CenterOfMassFrame}
\end{align}
and which directly implies $\mathbf{\|p\|}= \mathbf{\|l\|}$. 
To sum up, the momentum variables we are left with are:
\begin{enumerate}
\item[(i)] An independent variable $p_0$, which is what we wish to continue in the complex plane.
\item[(ii)] An independent three-momentum $\mathbf p$, which is a  spectator and acts as an external parameter.
\item[(iii)] Three angles $\phi_{\text{in}},\phi_{\text{out}},\theta_{\text{sc}}$, which are parameters that do not affect the Wick rotation. Of these, the first two are fixed by \labelcref{eq:CenterOfMassFrame}.
\end{enumerate}
Thus, we see that we have reduced the problem of an ill-defined set of momentum variables to the familiar case of a single frequency $p_0$ to be analytically continued. 
The trade-off is that the continuation has to be performed for different values of the parameters $\mathbf p,\theta_{\text{sc}}$. 
Only then, the on-shell limit $p_0 = - i \|\mathbf p\|$ can be taken. 
In the end, momentum conservation combined with Lorentz invariance and the on-shell condition, leaves us with two independent variables $p_0,\theta_{\text{sc}}$, or equivalently the Mandelstam variables $s,t$.

\subsection{Vertex Reconstruction}
\label{sec:ResultsForAmplitude}

As pointed out, the challenging task is the Wick rotation of the quantum part, namely the vertex dressing $\sqrt{G_{\phi\phi h} ( p_{ \text{in} }, z_{ \text{in} } ) \, G_{\phi\phi h} (p_{ \text{out} },z_{ \text{out} }) }$, as this can only be computed numerically and encodes the fixed-point information. Let us define the following object, a vertex function of $p_0$, 
\begin{align}
    V(\|\mathbf p\|; p_0) = \sqrt{G_{\phi\phi h}(p_{ \text{in} }, z_{ \text{in} }) \, G_{\phi\phi h}(p_{ \text{out} },z_{ \text{out} })}\,,
\label{eq:VertexReconstructing}
\end{align}
where $p_{\text{in}}, z_{\text{in}}, p_{\text{out}}, z_{\text{out}}$ are related to $p_0$ as per \labelcref{eq:NewMomentumVariables}. In essence, the Euclidean amplitude scales in the frequency domain as
\begin{align}
    \mathcal{A}_s = p_0^2\, V(\|\mathbf p\|; p_0)\, ,
\label{eq:AmplitdueAndP0Vertex}
\end{align}
and this is shown in \Cref{fig:VertexExternalP} for increasing values of $\|\mathbf p\|=\|\mathbf l\|$. We see that the vertex dressing inherits the asymptotic safety properties of $G_{\phi\phi h}$, see \Cref{fig:FullAngularCoupling}: for $p_0\to\infty$, $V(\|\mathbf p\|; p_0)$ scales as $1/p_0^2$, which in turn implies $\mathcal{A}_s \to $ constant. This is another hallmark of asymptotic safety. In particular, this means that the vertex dressing suppresses the amplitude at high energies. 

We also note the appearance of further structures at intermediate momenta $p_0\approx 7\,M_{\text{pl}}$ in \Cref{fig:VertexExternalP}. In particular, as $\|\mathbf p\|$ increases, a peak-like structure appears. 
We highlight that the Newton coupling displayed in \Cref{fig:FullAngularCoupling} is monotonically decreasing in the $p$ direction for fixed $z$, while the converse is not true: a local maximum is present in the $z$ direction, see also \Cref{fig:z-dep-fixed-point}. The appearance of a maximum in $V(\|\mathbf p\|; p_0)$ is a consequence of the change of variables \labelcref{eq:NewMomentumVariables}, which mixes the $p$- and $z$-dependences. While the physics nature is unclear, this complicates the Wick rotation as peaks can lead to branch cuts and singularities in the complex plane. We stress that the object shown in \Cref{fig:VertexExternalP} is off-shell, while the observable amplitude will be obtained in the on-shell limit. In what follows, we illustrate the reconstruction procedure used to Wick rotate \Cref{fig:VertexExternalP} and \labelcref{eq:VertexReconstructing} to arrive at  the result already displayed in \Cref{fig:AmplitudeResult}. The reader interested in the physics consequences can jump to \Cref{sec:LorentzianAmplitude}.

The Wick rotation of the vertex function to the Lorentzian branch is done by means of a reconstruction procedure. For this, we assume that the vertex \labelcref{eq:VertexReconstructing} possesses a K\"all\'en-Lehmann representation, namely, it can be expressed as
\begin{align}
    V(\|\mathbf p\|; p_0) = \int_0^\infty \frac{\mathrm{d} \lambda \,\lambda}{\pi}\ \frac{\rho(\|\mathbf p\|; \lambda)}{\lambda^2 + p_0^2}\,.
\label{eq:KLRepresentation}
\end{align}
From this, the spectral function $\rho(\|\mathbf p\|; \lambda)$ can be obtained by 
\begin{align}
    \rho(\|\mathbf p\|; \lambda) = 2\ \Im V(\|\mathbf p\|; -\mathrm{i}(\lambda + \mathrm{i}0^+) )  \,.
\label{eq:SpectralFunction}
\end{align}  
We do not discuss any property of $\rho(\|\mathbf p\|;\lambda)$, such as positivity or normalisation, and we just use it to perform the Wick rotation.

Once $\rho$ is known, we can rotate $p_0\to -i \|\mathbf p\|$, thus finally obtaining the on-shell Lorentzian vertex dressing
\begin{align}
    V_{L}(\|\mathbf p\|) = \int_0^\infty \frac{\mathrm{d} \lambda \,\lambda}{\pi}\ \frac{\rho(\|\mathbf p\|;\lambda)}{\lambda^2 - \|\mathbf p\|^2 + i0^+}\,.
\label{eq:LorentzianVertexDressing}
\end{align}
A full reconstruction has proved surprisingly difficult for a number of reasons. First, the non-trivial peak in \Cref{fig:VertexExternalP} complicates the reconstruction, as rational approximants, such as Padé-type methods, tend to introduce spurious poles in the upper-half plane. Second, more refined approaches such as BW fits or GPR provide successful reconstructions, but suffer from well-known overfitting issues, see \Cref{app:comparisonRecMethods}.

\begin{figure*}[tbp]
    \includegraphics[width=0.48\textwidth]{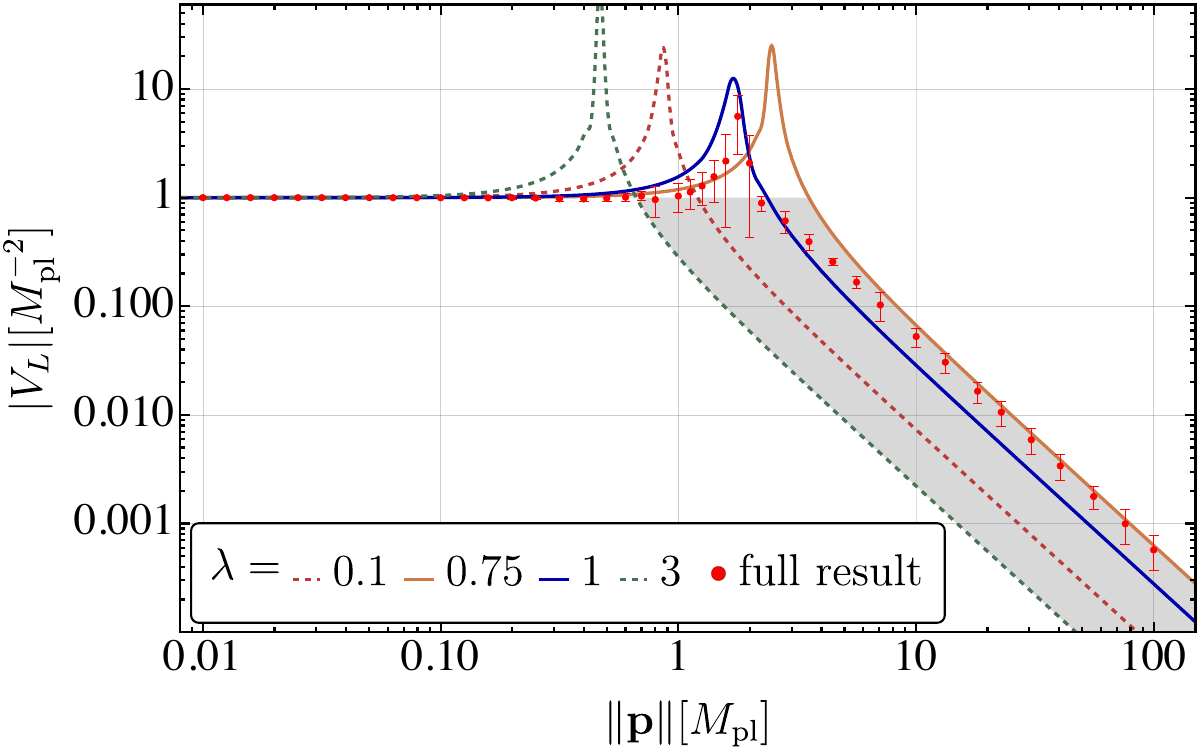}\hfill
    \includegraphics[width=0.465\textwidth]{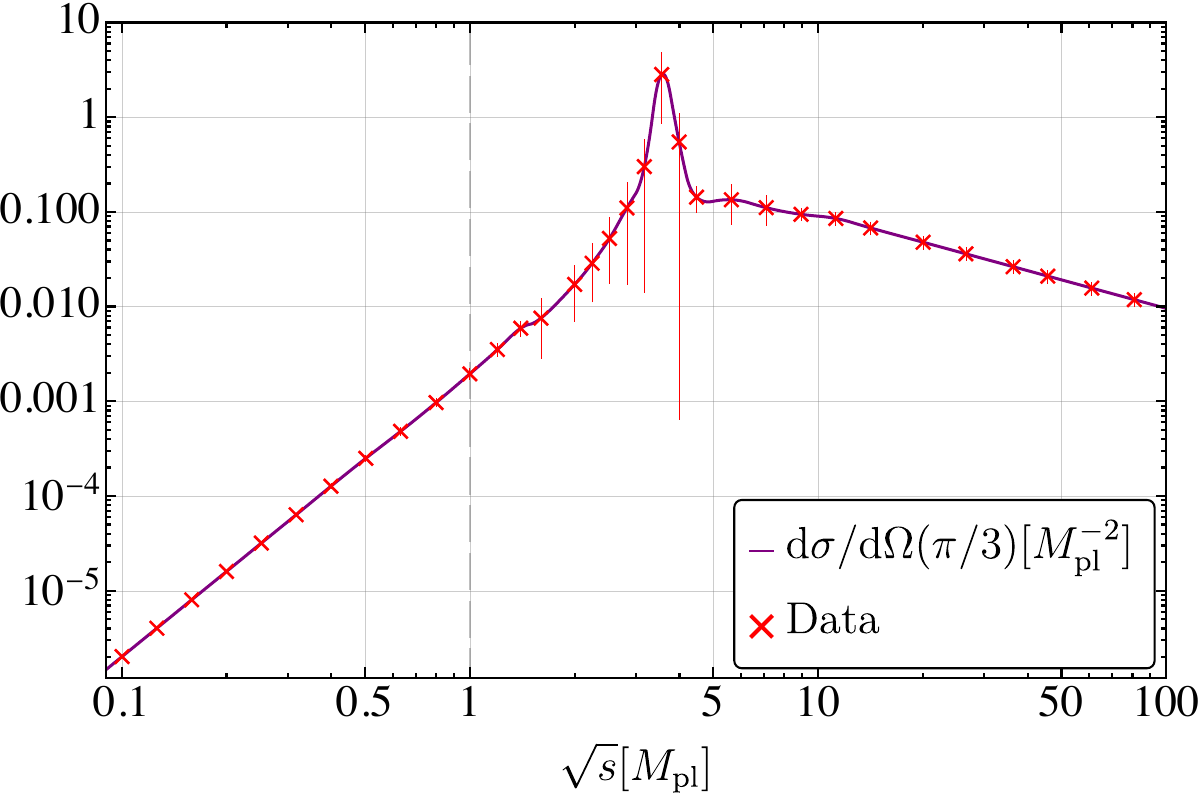}
    \caption{Left: Comparison of the Lorentzian vertex $V_L(p_0)$ for different reconstruction methods. The red data represents the full off-shell reconstruction of \Cref{sec:ResultsForAmplitude}, which is used to compute \Cref{fig:AmplitudeResult} and $\mathrm d \sigma/\mathrm d \Omega$ in the right panel. The solid and dotted curves correspond to the simpler on-shell reconstruction for different values of $\lambda$, see \Cref{sec:ComparisonWithSimplfiedReconstruction}. The qualitative features are shared across the different reconstructions. Right: Differential cross section $\mathrm d\sigma/\mathrm d\Omega$ as defined in \labelcref{eq:DifferentialCrossSection} as a function of the centre of mass energy $\sqrt{s}$ for a fixed scattering angle $\theta_{\text{sc}}=\pi/3$. We display the computed data (red) together with a smooth interpolation (purple). The dashed grey line indicates the classical Planck scale.}
    \label{fig:AmplitudeAndCrossSection}
\end{figure*}

While a full off-shell reconstruction is desirable, it is not necessary as long as one reconstructs sufficiently well in the neighbourhood of the on-shell point $p_0 = - i \|\mathbf p\|$. Following \cite{Binosi:2019ecz}, we approximate each off-shell vertex $V(\|\mathbf p\|; p_0)$ via a continued fraction $C_N(\|\mathbf p\|; p_0)$ of order $N$, see also \Cref{app:SPM}. Such a rational method introduces spurious poles in the complex plane, but it can be improved by finding the optimal set of points such that the original data is well reconstructed.

We devise the following algorithm:
\begin{enumerate}
\item[(1)] Let $M$ be the total number of points in the original data set at fixed $\|\mathbf p\|$,
\begin{align}
    S &= \left\{ \left( p^{(i)}_0\,, V^{(i)} \right) \right\}\,,& i&=1,\dots,M\,,
\label{eq:MDataset}
\end{align}
and fix an order $N<M$ for the continued fraction $C_N(p_0)$. We take $N=80$.
\item[(2)] Randomly select $N$ points from the set $S$:
\begin{align}
    S \supset S' &= \left\{ \left( p^{(j)}_0\,, V^{(j)} \right) \right\}\,.
\label{eq:NDataset}    
\end{align}
\item[(3)] Compute the continued fraction $C_N(p_0)$ on the set $S'$.
\item[(4)] Obtain the spectral function $\rho_M(\lambda)$ via \labelcref{eq:SpectralFunction}, and the reconstructed vertex $V_{\text{rec},N}(p_0)$ via \labelcref{eq:KLRepresentation}.
\item[(5)] Compute the $\chi^2$ error between the original data and the reconstructed one on the full set S:
\begin{align}
    \chi^2 = \frac{1}{M} \sum_{i=1}^M \left[ V\!\left(\|\mathbf p\|; p_0^{(i)}\right) - V_{\text{rec},N}\!\left(\|\mathbf p\|;p_0^{(i)}\right) \right]^2.
\label{eq:chi2}
\end{align}
\item[(6)] Fix a threshold for the $\chi^2$ test. In our case, we fix $\chi^2_{T}=20$ for the $\|\mathbf p\|$ tails and $\chi^2_{T}=5$ for the intermediate region $\|\mathbf p\|\approx 1$. This is justified by the fact that in the asymptotics, the vertex is smooth and easier to reconstruct, while in the intermediate region the peak complicates the reconstruction.
\item[(7)] If $\chi^2>\chi^2_T$, then select another random set $S''\subset S$ of $N$ points, and repeat steps (3) through (6) until $\chi^2 < \chi^2_T$.
\item[(8)] Once one optimal set of points is found, compute the on-shell Lorentzian vertex $V_L(\|\mathbf p\|)$ \labelcref{eq:LorentzianVertexDressing}.    
\item[(9)] Iterate steps (2) through (8) for $K$ times and average the results. In our case, we take $K=80$.
\item[(10)] Repeat steps (1) through (9) for every value of $\|\mathbf p\|$.
\end{enumerate}
At the end of the algorithm, we are left with an on-shell Lorentzian vertex dressing $V_L(\|\mathbf p\|)$, which can then be plugged into \labelcref{eq:AmplitudeContraction} to obtain the full Lorentzian amplitude. In the left panel of \Cref{fig:AmplitudeAndCrossSection}, we show the reconstruction result for different values of $\|\mathbf p\|$ in red. We obtain a statistical error from the reconstruction, which is displayed in \Cref{fig:AmplitudeResult,fig:AmplitudeAndCrossSection}. We do not take into account errors from approximations of the flow equation.

\section{Scattering amplitude}
\label{sec:LorentzianAmplitude}

\subsection{Lorentzian Results}
\label{subsec:LorentzianResults}

The non-perturbative, fully resummed, amplitude in ASQG is shown in \Cref{fig:AmplitudeResult} for the $s$-channel.
$\mathcal{A}_{s}$ was defined in \labelcref{eq:AmplitudeContraction}, and in terms of Mandelstam variables it explicitly reads
\begin{align}
    \mathcal{A}_s = V_L(s)\, \frac{1}{s} \left( t^2 -4tu +u^2 - s^2 \right),
\label{eq:LorentzAmplitudeMandelstam}
\end{align}
where the subscript indicates we are taking the Lorentzian version of the vertex dressing as reconstructed in \Cref{sec:ResultsForAmplitude}. In the centre of mass frame \labelcref{eq:CenterOfMassFrame}, this reduces to
\begin{align}
    \mathcal{A}_{s} = - s\, \sin^2( \theta_{\text{sc}} )V_L(s)\,,
\label{eq:LorentzianAmplitude}
\end{align}
in terms of the centre-of-mass energy $\sqrt{s}$ and the scattering angle $\theta_{\text{sc}}$, and it is displayed in \Cref{fig:AmplitudeResult} for a given value of $\theta_{\text{sc}}$. Once the amplitude is known, the differential cross section is straightforwardly obtained via 
\begin{align}
    \frac{\mathrm d\sigma_s}{\mathrm d\Omega} = \frac{1}{64\pi^2 s}\ |\mathcal{A}_s|^2\,.
\label{eq:DifferentialCrossSection}
\end{align}
Note that we are projecting on the $s$-channel of \Cref{fig:scattering} only. From this, the total cross section can be obtained by integrating over the solid angle $\mathrm{d}\Omega$. This yields
\begin{align}
    \sigma_s = \frac{27}{64\pi} s |V_L(s)|^2\,.
\label{eq:IntegratedCrossSection}
\end{align}
The result for the differential cross section is displayed in the right panel of \Cref{fig:AmplitudeAndCrossSection}.

Below the Planck scale, quantum gravity fluctuations are suppressed and the amplitude and cross-sections agree with classical GR since the physical Newton coupling remains classical in the IR limit with $G_{\phi\phi h}\to G_\text{N}$, see \labelcref{eq:ClassicalNewtonCoupling}. In this regime, the amplitude grows linearly with $s$, and likewise the cross section. Above the Planck scale, the fixed-point regime dominates, which is rooted in the running of $G_{\phi\phi h}$ with the physical momentum $p$. In the fixed-point regime, the cross section decreases as $1/s$ for $s\gg M_\text{pl}^2$. The momentum-dependence of the vertex dressing makes the cross section decay and the amplitude bounded. The latter satisfies the Froissart bound \cite{Froissart:1961ux} and both are compatible with requirements from unitarity. 

The UV and IR regimes are connected by a peak appearing at $\sqrt{s}\simeq 5 M_\text{pl}$. This peak is absent in the Euclidean data at fixed $z$, see \Cref{fig:FullAngularCoupling}, but a local maximum does appear in the off-shell vertex, see \Cref{fig:VertexExternalP}. Such a peak in the amplitude was also observed in \cite{Pastor-Gutierrez:2024sbt}. The resonance might be interpreted as a graviton bound state but more detailed studies are required to understand the resonant behaviour. Evidence in favour of graviton bound states have also been observed on the lattice \cite{Maas:2025rug, Maas:2025mne}.

\subsection{Comparison with the Simplified Reconstruction}
\label{sec:ComparisonWithSimplfiedReconstruction}

We cross-check the algorithm employed in \Cref{sec:ResultsForAmplitude} with a comparison to a different reconstruction procedure. We fix the momentum variables as $p_0=\lambda \|\mathbf p\|$, with $\lambda\in[0,\infty)$, which allows us to scan the stability of the reconstruction around the on-shell condition on the Euclidean branch. We are effectively reducing the problem to a single momentum variable $\|\mathbf p\|$. Inserting the $\lambda$ parameterisation in \labelcref{eq:NewMomentumVariables}, we get 
\begin{align}
    p^2&=(1+\lambda^2)\|\mathbf p\|^2\,,
    &
    z&= \frac{ -1+\lambda^2 }{ 1+\lambda^2}\,,
\label{eq:LambdaAndpAndz}
\end{align}
where we have also fixed the centre of mass frame \labelcref{eq:CenterOfMassFrame}. A consequence of the approximation is that the amplitude is only sensitive to one momentum configuration of the vertex dressing. In consequence, the vertex dressing is a rather simple function and can be reconstructed with a simple Padé. In Padé, we implement the IR behaviour $V(p_0)=1$ and the UV asymptotic behaviour $V(p_0)\sim p_0^{-2}$, to wit
\begin{align}
    \mathcal{P}[N/N+2](p_0) = \dfrac{ 1 + \sum_{n=1}^{N} a_n p_0^n}{1 + \sum_{n=1}^{N} b_{n+2} p_0^{n+2} }\,.
\label{eq:PadeApproximant}
\end{align}
The coefficients $a_n,b_n$ are determined by a fit to the numerical data. For our case, $N=3$ is sufficient to obtain a converged result. The Lorentzian vertex is obtained at each lambda via
\begin{align}
    V_{L,\lambda}(p_0) = \mathcal{P}[3/5](i p_0)\,.
\end{align}
In the left panel of \Cref{fig:AmplitudeAndCrossSection}, we display the vertex $V_L$ for different values of $\lambda$ from $0.1$ to $3$, where we highlight the different UV tails with a grey band. Interestingly, the UV tail of $V_L$ has a maximum at $\lambda = 0.75$. Inserting this value in \labelcref{eq:LambdaAndpAndz}, we find that it corresponds to the angle $z\approx-0.28$, which is the maximum of $g^*(z)$, see \Cref{fig:z-dep-fixed-point}. Overall we find a good qualitative agreement of the simplified reconstruction with the reconstruction employed in \Cref{sec:ResultsForAmplitude}. Remarkably, the latter lies almost at the upper bound of the UV band of the simplified reconstruction.

\begin{figure}[tbp]
    \includegraphics[width=\columnwidth]{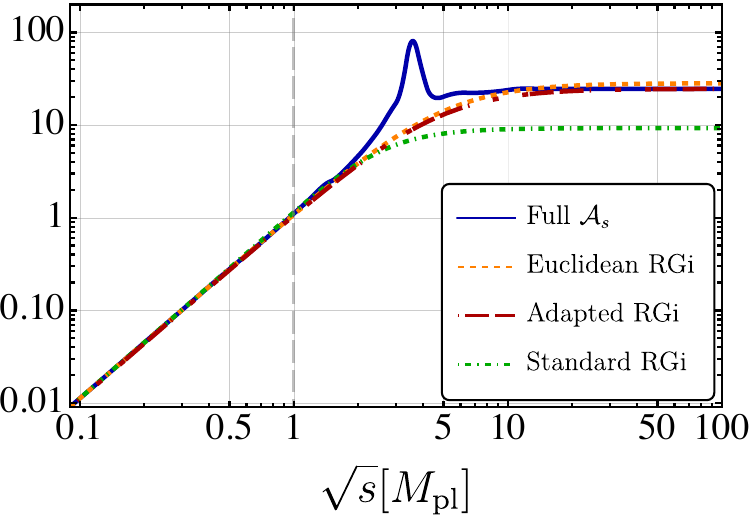}
    \caption{The $s-$channel amplitude for the process $\phi\phi\to h \to \phi\phi$. We depict in blue the full Lorentzian result of \Cref{fig:AmplitudeResult}. The different dashed lines represent different RG improvement schemes. In green, we identify the cutoff $k$ with $\sqrt{s}$ in the momentum-independent coupling $G(k,0,z=-0.28)$. In orange, we employ the Euclidean vertex $G(0,p,z=-0.28)$ of \Cref{fig:FullAngularCoupling}, without performing any Wick rotation, and we identify $p=\sqrt{s}$. Last, in red, we employ the adapted RGi procedure \labelcref{eq:AdaptedRGiTrajectory}, which takes into account the fixed-point value of the Lorentzian vertex.}
    \label{fig:AmplitudeComparisonRGI}
\end{figure}

\subsection{Comparison to RG Improvement}
\label{sec:ComparisonWithRGImprovement}

A convenient way to include quantum effects into observables is to identify the cutoff scale $k$ with one physical scale of the process. For the case of an amplitude like \labelcref{eq:LorentzianAmplitude}, the natural choice is to identify $k$ with the centre-of-mass energy, $k^2=s$. This procedure goes under the name of RG improvement. In general, it is expected to be a decent approximation for leading-order effects when the system has only one physical scale. The latter holds if we focus on one single diagram, for example, the $s$-channel mediated diagram.

We express the avatar of the Newton coupling as a function of the RG scale $k$, the Euclidean momentum $p$ and the angle $z$,
\begin{align}
    G(k,p,z)\,.
\label{eq:VertexEasyNCoupling}
\end{align}
We fix the angle $z$ such that the UV behaviour of the Lorentzian vertex $V_L$ is best captured, which corresponds to $z=-0.28$. Now, we follow \cite{Pastor-Gutierrez:2024sbt} and define several types of RG improvements:
\begin{enumerate}
\item \textit{Standard RG improvement}: We identify $k^2=s$ and replace the vertex dressing in \labelcref{eq:LorentzianAmplitude} with the cutoff dependent coupling,
\begin{align}
V_L(s) \longrightarrow G(\sqrt{s}, 0, z=-0.28)\,.
\label{eq:StandardRGi}
\end{align}
This is depicted with a green dashed-dotted line in \Cref{fig:AmplitudeComparisonRGI}.
\item \textit{Euclidean RG improvement}: We identify the Euclidean momentum with the centre-of-mass energy $p^2=s$ and replace the vertex dressing $V_L(s)$ in \labelcref{eq:LorentzianAmplitude} with the Euclidean coupling at $k\to0$, (shown also in \Cref{fig:FullAngularCoupling}), without performing any Wick rotation,
\begin{align}
V_L(s) \longrightarrow G(0, \sqrt{s}, z=-0.28)\,.
\label{eq:EuclideanRGi}
\end{align}
This is depicted with an orange dashed line in \Cref{fig:AmplitudeComparisonRGI}.
\item \textit{Adapted RG improvement}: From \Cref{fig:AmplitudeAndCrossSection} we see that the on-shell Lorentzian vertex itself possesses a fixed point, $V^*$. We can then adapt the RG improving procedure to be sensitive to this fixed-point value and define the trajectory,
\begin{align}
V_{\textrm{RG}}(s) = \frac{V^*}{s + V^* M_{\text{pl}}^2}\,,
\label{eq:AdaptedRGiTrajectory}
\end{align}
and substitute the vertex dressing in \labelcref{eq:LorentzianAmplitude} with
\begin{align}
V_L(s) \longrightarrow V_{\textrm{RG}}(s)\,.
\label{eq:AdaptedRGiSubstitution}
\end{align}
This is depicted with a red dashed line in \Cref{fig:AmplitudeComparisonRGI}.  %
\end{enumerate}
In \Cref{fig:AmplitudeComparisonRGI} we see that every RG improvement captures the correct UV scaling of the amplitude. The standard RG improvement procedure gets the value wrong by almost an order of magnitude, since the RG fixed point at $p=0$ is smaller than the momentum fixed point at $k=0$, see \Cref{fig:z-dep-fixed-point}. On the other hand, the Euclidean and the adapted scheme agree reasonably well with the full result in the deep UV, the latter by construction. While the UV is reasonably well captured, every RG improvement fails to capture the peak structure around the Planck scale.

\subsection{Forward Scattering}
\label{sec:BeyondS-channel}

So far we have focused on the analysis of the $s$-channel. The $t$- and $u$-channels are obtained straight-forward by crossing symmetry. The structure is fully analogous, and, for example, the $t$-channel amplitude reads
\begin{align}
\mathcal{A}_{t} = V_L(t)\, \frac{1}{t} \left( s^2 -4su +u^2 -t^2 \right) \,.
\label{eq:LorentzianAmplitudeT-channel}
\end{align}
Here, a crucial difference between the $s$ and $t$ channel shows up: The tree-level amplitude contains an $\sim s^2/t$ divergence and since the vertex function depends only on $t$, the amplitude is still divergent in the non-perturbative computation, which is a well known issue of the forward limit where $t$ is fixed and $s$ goes to infinity. The divergence in the forward limit can only be resolved with the inclusion of $\mathcal{A}_4$, see also \cite{Draper:2020bop}.

\section{Conclusions}
\label{sec:conclusions}
In this work, we presented the first non-perturbative, resummed scattering amplitude for the process of four identical scalar fields interacting via graviton exchange, shown in \Cref{fig:AmplitudeResult}. The amplitude follows the behaviour expected from general relativity in the infrared (IR), while approaching a constant in the ultraviolet (UV) once the scale-invariant regime of asymptotic safety is reached. The resulting amplitude therefore remains bounded at high energies and is compatible with unitarity.

This result was obtained by computing the full momentum dependence of the renormalisation group flow of the scalar–gravity vertex $\Gamma_{k}^{\phi\phi h}$. Solving this flow allowed us to integrate out all quantum fluctuations and determine the physical 1PI vertex of the quantum effective action at $k=0$. The resulting vertex exhibits a clear signature of a physical realisation of asymptotic safety: it approaches a non-trivial UV fixed point characterised by a $p^{-2}$ scaling of the vertex dressing at large momenta. Importantly, this scaling behaviour holds across all momentum configurations. This demonstrates explicitly how the RG fixed point of the Newton coupling in the $k$-direction translates into a physical momentum scaling of correlation functions in the UV, including those depending on multiple momenta.

The Euclidean vertex enters directly into the graviton-mediated scalar–scalar scattering amplitude, more precisely into the off-shell analytic continuation of the amplitude. To obtain the physical Lorentzian observable, a Wick rotation followed by the on-shell limit is required. We performed the Wick rotation using a reconstruction procedure, which is not unique and may in principle introduce artefacts. To assess the robustness of our results, we therefore implemented several independent reconstruction methods, including a simplified procedure that applies the on-shell condition prior to the Wick rotation. All reconstruction approaches yield qualitatively consistent results, as shown in \Cref{fig:AmplitudeAndCrossSection}.

Our main results are the Lorentzian scattering amplitude and the associated cross section. In the IR, both observables reproduce the behaviour expected from general relativity. In the UV, the fixed-point properties of the Newton coupling leave a clear imprint on the physical observables: the non-perturbative asymptotically safe scaling becomes directly visible in the scattering amplitude, which approaches a constant at high energies. Consequently, the amplitude remains bounded and compatible with unitarity. In the transition region between the IR and UV regimes, we observe a resonance-like structure. This feature may indicate the presence of a physical resonance, potentially associated with a graviton bound state. Clarifying the origin of this structure will require further investigation.

\begin{acknowledgments}
We thank Benjamin Knorr for discussions. JMP is funded by the Deutsche Forschungsgemeinschaft (DFG, German Research Foundation) under Germany’s Excellence Strategy EXC 2181/1 - 390900948 (the Heidelberg STRUCTURES Excellence Cluster) and the Collaborative Research Centre SFB 1225 - 273811115 (ISOQUANT). MR acknowledges support by the Science and Technology Facilities Council under the Consolidated Grant ST/X000796/1 and the Ernest Rutherford Fellowship ST/Z510282/1.
\end{acknowledgments}

\appendix

\section{Gauge Fixing}
\label{app:GaugeFixing}
The EH action \labelcref{eq:EHAction} is augmented by a gauge-fixing action,
\begin{align}
    S_\text{gf}[\bar g;h]&=\frac{1}{32 \pi G_\text{N} \alpha}\int \! \mathrm d^4x\sqrt{\bar g}\, F_{\mu}\bar g^{\mu\nu}F_\nu\,, \notag\\[1ex]
    F_\mu&=\bar\nabla^\lambda h_{\mu\lambda}-\frac{1+\beta}{4}\bar\nabla_\mu h^\lambda_\lambda\,,
\label{eq:GfAction}
\end{align}
together with a corresponding Faddeev-Popov ghost action,
\begin{align}
    S_\text{gh}[\bar g;&h,c,\bar c]=\int \! \mathrm d^4x\sqrt{\bar g}\ \bar c^\mu\mathcal{M}_{\mu\nu}c^\nu\,, \notag\\[1ex]
    \mathcal{M}_{\mu\nu} &= \bar\nabla^\lambda(g_{\mu\nu}\nabla_\lambda-g_{\lambda\nu}\nabla_\mu)+\frac{1+\beta}{2}\bar g^{\rho\lambda}\bar\nabla_\mu g_{\nu\rho}\nabla_\lambda\,.
\label{eq:GhostAction}
\end{align}
We work in the Landau-limit of the harmonic gauge, given by
\begin{align}
    \alpha&=0\,,
    &
    \beta&=1\,.
\label{eq:HarmonicGauge}
\end{align}
%

\section{Vertex Expansion}
\label{app:VertexExpansion}

In this work, we use a vertex expansion of the scale-dependent effective action. Using the notation of correlation functions $\Gamma^{(n)}_k(p_1,...,p_n)$ as given in \labelcref{eq:CorrelationFunctions}, the scale-dependent action can be written as
\begin{align}
    \Gamma_k[\Phi] = \sum_{n=0}^\infty \frac{1}{n!}\int_{p_1,...,p_n} \!\!\Gamma^{(n)}(p_1,...,p_n)\ \Phi_1(p_1)\cdots\Phi_n(p_n)\,.
\label{eq:VertexExpansion}
\end{align}
We parametrise the correlation functions as
\begin{align}
    \Gamma^{(n)}_{k}(p_1,...,p_n) = \sum_j\ \gamma^{(n)}_{j,k}(p_1,...,p_n)\ \mathcal{T}_j^{(n)}(p_1,...,p_n),
\label{eq:CorrelationFunctionParametrization}
\end{align}
where the $\mathcal{T}_j^{\Phi_1\dots\Phi_n}(p_1,...,p_n)$ are the tensor structures associated with the given vertex, and the $\gamma^{(n)}_{j,k}(p_1,...,p_n)$ are the corresponding vertex dressings or couplings. The sum over $j$ is a sum over a complete basis of tensor structures. In general, the number of tensor structures increases with $n$, and thus the parametrisation of \labelcref{eq:CorrelationFunctionParametrization} becomes more and more involved.

In this work, we compute the correlation function $\Gamma^{\phi\phi h}_{k}(p_1,p_2)$. This vertex is symmetric under the exchange of momenta, $p_1 \leftrightarrow p_2$, and indices, $\mu \leftrightarrow \nu$. Its basis contains three elements,
\begin{align}
    \Gamma^{\phi\phi h}_{k}(p_1,p_2) = \sum_{j=1}^3 \gamma^{\phi\phi h}_{j,k}(p_1,p_2)\ \mathcal{T}_j^{\phi\phi h}(p_1,p_2) + (p_1 \leftrightarrow p_2)\,.
\label{eq:VertexTensorBasisSum}
\end{align}
A possible choice of basis is given by
\begin{align}
    \mathcal{T}^{\phi\phi h}_1 & = \delta_{\mu\nu}\,,
    &
    \mathcal{T}^{\phi\phi h}_2 & = \frac{p_1^{(\mu} p_2^{\nu)}}{|p_1||p_2|}\,,
    &
    \mathcal{T}^{\phi\phi h}_3 & = \frac{p_1^{\mu} p_1^{\nu}}{p_1^2}\, .  
\label{eq:ScalarTensorBasis}
\end{align}
The tensor basis is also spanned by the three independent curvature invariants entering the $\phi\phi h$ vertex, namely $\sqrt{g}\phi^2$, $\sqrt{g} R \phi^2$, and $\sqrt{g} R_{\mu\nu} \phi \partial_\mu \partial_\nu \phi$, see also \cite{Knorr:2022dsx}, and there is a bijective map between the tensor structures of the curvature invariants and \labelcref{eq:ScalarTensorBasis}.

For most graviton correlation functions, it is technically too challenging to implement a complete basis. Instead, one isolates a tensor structure of interest with a suitable projection. Here, we use a projection on the classical tensor structure. For the scalar action \labelcref{eq:ScalarAction}, the classical tensor structure of the three-point function reads 
\begin{align}
    \mathcal{T}^{\phi\phi h}_\text{classical}(p_1,p_2) &= \frac{\delta^3 S_\phi } { \delta h_{\mu\nu}(p_h)\, \delta\phi(p_1)\, \delta\phi(p_2) } \bigg\rvert_{ h=\phi=0,\bar g=\delta } \notag\\
    &= \frac{1}{2} \left( \delta^{\mu\nu} p_1 \cdot p_2 - p_1^{(\mu} p_2^{\nu)} \right).
\label{eq:ScalarTensorStructure}
\end{align}
%

\section{Graviton Propagator}\label{app:GravitonPropagator}
A convenient decomposition of the fluctuation field $h_{\mu\nu}$ is made of a spin-2 mode $h^{tt}_{\mu\nu}$, a spin-1 mode $\xi_\mu$ and two spin-0 modes. The decomposition is then reflected at the level of the propagator. On a flat background, the decomposition can be implemented with the Stelle spin projection operators. We define the transverse and longitudinal projection operators as
\begin{align}
    \Pi^t_{\mu\nu}&=\delta_{\mu\nu}-\frac{p_\mu p_\nu}{p^2}\,,
    &
    \Pi^l_{\mu\nu}&=\frac{p_\mu p_\nu}{p^2}\,.
\label{eq:TransverseLongitudinalProjectors}
\end{align}  
The spin projections are defined as
\begin{align}
    \Pi^{tt}_{\mu\nu\rho\sigma} &= \frac{1}{2}\left(\Pi^t_{\mu\rho}\Pi^t_{\nu\sigma}+\Pi^t_{\nu\rho}\Pi^t_{\mu\sigma}\right) - \frac{1}{3}\Pi^t_{\mu\nu}\Pi^t_{\rho\sigma}\,,\notag\\[1ex]
    \Pi^{v}_{\mu\nu\rho\sigma} &= \frac{1}{2}\left(\Pi^t_{\mu\rho}\Pi^l_{\nu\sigma}+\Pi^t_{\nu\rho}\Pi^l_{\mu\sigma}+\Pi^t_{\mu\sigma}\Pi^{l}_{\nu\rho}+\Pi^t_{\nu\sigma}\Pi^l_{\mu\rho}\right),\notag\\[1ex]
    \Pi^{0,s}_{\mu\nu\rho\sigma} &= \frac{1}{3}\Pi^t_{\mu\nu}\Pi^t_{\rho\sigma}\,,\notag\\[1ex]
    \Pi^{0,w}_{\mu\nu\rho\sigma} &= \Pi^l_{\mu\nu}\Pi^l_{\rho\sigma}\,,\notag\\[1ex]
    \Pi^{0,sw}_{\mu\nu\rho\sigma} &= \Pi^{0,ws}_{\rho\sigma\mu\nu} = \frac{1}{\sqrt{3}}\Pi^t_{\mu\nu}\Pi^l_{\rho\sigma}\,.
\label{eq:SpinProjectors}
\end{align}
The propagator is then decomposed as
\begin{align}
    \mathcal{G}_{\mu\nu\rho\sigma}(p) = \sum_{i} \frac{32\pi}{Z_i(p)}\ \mathcal{G}^{(i)}(p)\ \Pi^{(i)}_{\mu\nu\rho\sigma}(p)\,, 
\label{eq:GravitonPropagatorDecomposition}
\end{align}
with the wave functions $Z_i$, one for each mode. Here, we work with an approximation where all $Z_i$ are taken to be equal. The scalar parts $ \mathcal{G}^{(i)}(p)$ read in the gauge \labelcref{eq:HarmonicGauge},
\begin{align}
    \mathcal{G}^{(tt)}(p)  &= \frac{1}{p^2 (1+r_h) + k^2\mu}\,, \notag \\[1ex]
    \mathcal{G}^{(0,s)}(p) &= -\frac{1}{2}  \frac{1}{p^2 (1+r_h) + k^2\mu}\,, \notag \\[1ex]
    \mathcal{G}^{(0,w)}(p) &= -\frac{3}{2}  \frac{1}{p^2 (1+r_h) + k^2\mu}\,, \notag \\[1ex]
    \mathcal{G}^{(0,sw)}(p) &= \mathcal{G}^{(0,ws)}(p) = -\frac{\sqrt{3}}{2}  \frac{1}{p^2 (1+r_h) + k^2\mu} \,.
\label{eq:GravitonPropagatoComponents}
\end{align}
For vanishing mass parameter $\mu=0$, the propagator can be written in a compact form as
\begin{align}
    \mathcal{G}_{\mu\nu\rho\sigma}(p) &= \frac{32\pi}{Z_h(p)}\ \frac{1}{p^2 (1+r_h) } \mathcal{P}_{\mu\nu\rho\sigma}\,,
\label{eq:GravitonPropagatorKto0}
\end{align}
where the tensor $\mathcal{P}_{\mu\nu\rho\sigma}$ is given by a combination of the Stelle projection operators,
\begin{align}
    \mathcal{P}_{\mu\nu\rho\sigma} &= \Pi^{tt}_{\mu\nu\rho\sigma} - \frac{1}{2} \Pi^{0,s}_{\mu\nu\rho\sigma} - \frac{3}{2}\Pi^{0,w}_{\mu\nu\rho\sigma} \notag \\[1ex]
    &\quad- \frac{\sqrt{3}}{2} \left( \Pi^{0,ws}_{\mu\nu\rho\sigma} + \Pi^{0,sw}_{\mu\nu\rho\sigma} \right)\,.
\label{eq:GravitonTensorStructureLandauG}
\end{align}
%

\section{Momentum Conventions}\label{app:MomentumConventions}
In this appendix, we summarise the momentum conventions used throughout this work. We call $p_{1,\mu}$ and $ p_{2,\mu}$ the external four-momenta belonging to the scalar legs, while the external momentum $ p_{h,\mu}$ belongs to the graviton leg. For the $O(4)$ angles between the momenta, we use
\begin{align}
	 p_{1,\mu}\ p_{2,\mu} &= \|p_1\| \|p_2\| z_{12}\,,\notag \\[1ex]
	 p_{1,\mu}\ p_{h,\mu} &= \|p_1\| \|p_h\| z_{1h}\,, \notag \\[1ex]
	 p_{2,\mu}\ p_{h,\mu} &= \|p_2\| \|p_h\| z_{2h}\,. 
\label{eq:ScalarProducts}
\end{align}
Momentum conservation leads to the relations
\begin{align}
	\|p_h\| z_{1h} &= -\|p_1\| - \|p_2\| z_{12}\,,\notag \\[1ex]
	\|p_h\| z_{2h} &= -\|p_2\| - \|p_1\| z_{12}\,,\notag \\[1ex]
    \|p_h\|^2 &= \|p_1\|^2 + \|p_2\|^2 + 2 \|p_1\| \|p_2\| z_{12} \,.
\label{eq:ScalarProdWConservation}
\end{align}
The vertex is fully parameterised by two momentum moduli $\|p_1\|,\|p_2\|$ and one Euclidean $O(4)$ angle $z_{12}=z$. 

For the loop integration $\int \mathrm d^4 q$, we use spherical coordinates with the conventions
\begin{align}
	 q_\mu &= q \begin{pmatrix}\cos(\theta_1) \\ \sin(\theta_1)\cos(\theta_2) \\ \sin(\theta_1)\sin(\theta_2)\cos(\phi) \\  \sin(\theta_1)\sin(\theta_2)\sin(\phi) \end{pmatrix} \notag \\[1ex]
     &= q \begin{pmatrix} x \\ \sqrt{1-x^2}\, y \\ \sqrt{1-x^2}\, \sqrt{1-y^2}\,\cos(\phi) \\  \sqrt{1-x^2}\, \sqrt{1-y^2}\,\sin(\phi) \end{pmatrix}\,,
\label{eq:def-spherical-coordinates}
\end{align}
where $\theta_i\in[0,\pi)$ and $\phi\in[0,2\pi)$. The integral measure is then given by
\begin{align}
	\mathrm d^4  q &= \mathrm d q\, \mathrm d \phi\, \mathrm d \theta_1\, \mathrm d \theta_2\, \sin(\theta_1)^2 \sin(\theta_2) q^3 \notag \\[1ex]
    &= \mathrm d q\, \mathrm d \phi\, \mathrm d y\, \mathrm d x \sqrt{1-x^2} \,q^3\,.
\end{align}
For the four-momenta, we parametrise the components as
\begin{align}
    p_{1,\mu} &= (\|p_1\|, 0, 0, 0)\,, \notag \\[1ex]
    p_{2,\mu} &= ( \|p_2\| \cos \varphi, \|p_2\| \sin \varphi, 0, 0)\,, \notag \\[1ex]
    p_{h,\mu} &= (-\|p_1\| - \|p_2\| \cos \varphi, - \|p_2\| \sin \varphi, 0, 0)\,,
\end{align}
with $\cos \varphi = z_{12}$. This implies for the scalar products
\begin{align}
	 q_\mu\ p_{1,\mu} &= q \,\|p_1\| x \,, \notag \\[1ex]
	 q_\mu\ p_{2,\mu} &= q\,  \|p_2\| (z_{12} x + \sqrt{1-z_{12}^2}\sqrt{1-x^2} \,y)\,,  \notag \\[1ex]
	 q_\mu\ p_{h,\mu} &= q (- \|p_1\| x - \|p_2\| z_{12} x - \|p_2\| \sqrt{1-z_{12}^2}\sqrt{1-x^2} \,y)  \,.
\end{align}
Lastly, Euclidean Mandelstam variables can be expressed as 
\begin{align}
	s&= ( p_{1,\mu}+  p_{2,\mu})^2 = \|p_1\|^2 + \|p_2\|^2 + 2 \|p_1\| \|p_2\| z_{12}\,, \notag\\[1ex]
	t&= ( p_{1,\mu}+  p_{3,\mu})^2 = \|p_1\|^2 + p_3^2 + 2 \|p_1\| p_3 z_{13}\,, \notag\\[1ex]
	u&= ( p_{1,\mu}+  p_{4,\mu})^2 = \|p_1\|^2 + p_4^2 + 2 \|p_1\| p_4 z_{14}\,.
\label{eq:MandelstamVariables}
\end{align}
%

\begin{figure*}[tbp]
    \centering
    \includegraphics[width=0.42\textwidth]{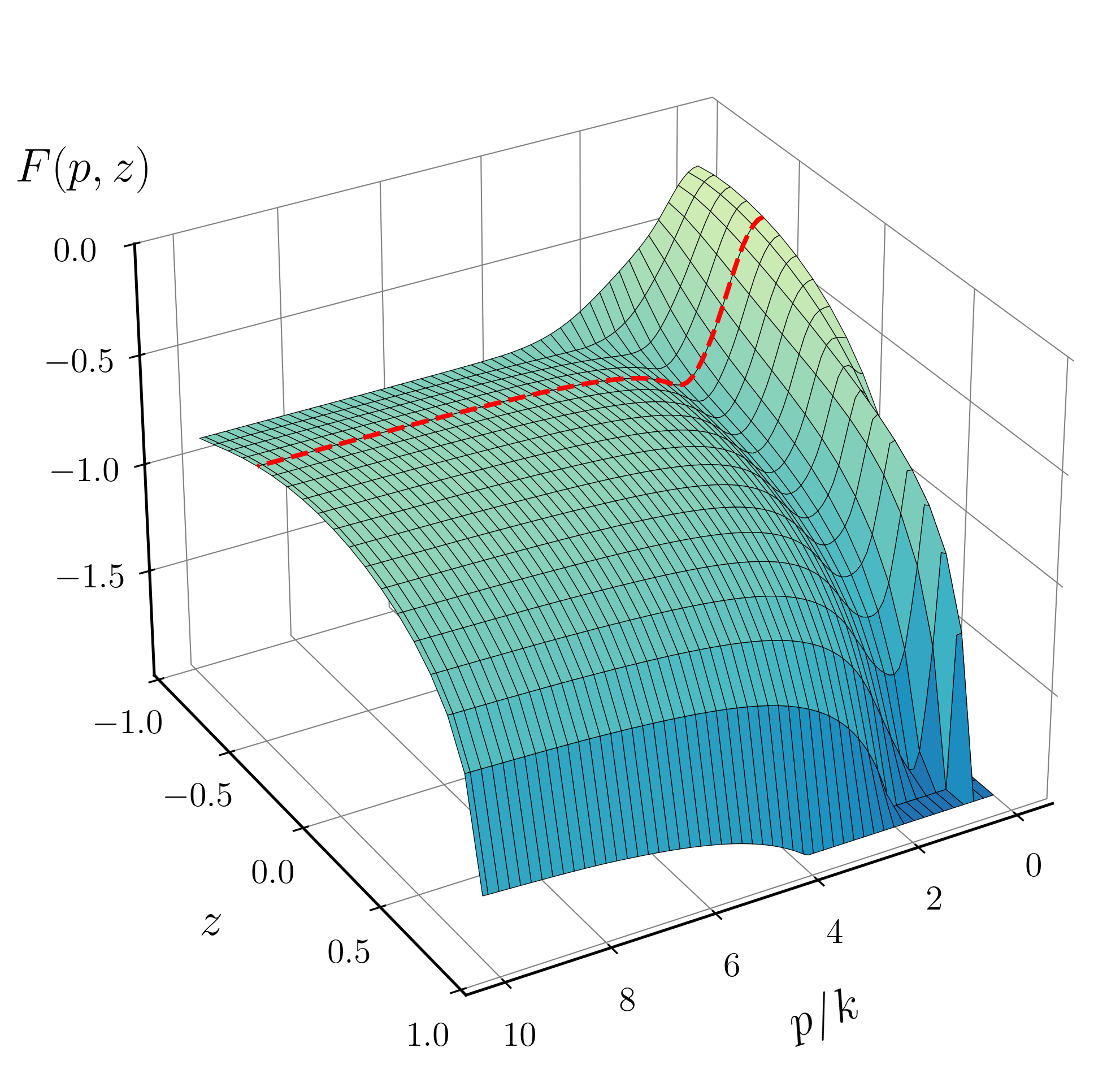}
    \hfill
    \includegraphics[width=0.47\textwidth]{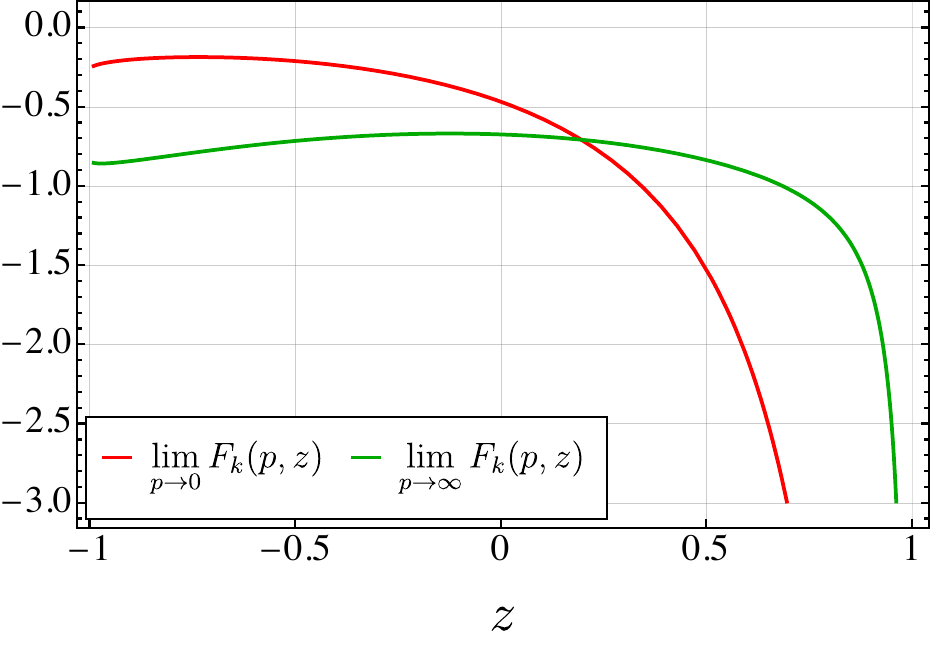}
    \caption{The flow function $F_k(p,z)$ in the transverse-traceless projection, \labelcref{eq:ttProjection}, left in the entire plane of the dimensionless momentum $p/k$ and the angle $z$, and right as a function of $z$ for $p=0$ and $p\to\infty$. The flow function diverges for $z\to1$.}
    \label{fig:FlowResolutionTT}
\end{figure*}

\section{Derivation of the Flow Equation}\label{app:FlowEquationDerivation}
In this appendix, we detail the derivation of \labelcref{eq:scalar-gravitonFlow}. We start from the flow of the three-point function $\Gamma^{\phi\phi h}_{k,\mu\nu}$, which we parametrise as given in \labelcref{eq:VertexParametrization}. The flow equation for the dimensionful coupling $G_{\phi\phi h}$ reads 
\begin{align}
    \frac{ \partial_t G_{\phi\phi h} }{2 \sqrt{ G_{\phi\phi h}} }\ \mathcal{ T }_{\mu\nu}^{ \phi\phi h } =
    \partial_t \bar{ \Gamma }^{\phi\phi h}_{\mu\nu}  - \left(\frac{1}{2} \eta_h + \eta_{\phi}\right) \sqrt{ G_{\phi\phi h} }\ \mathcal{T}_{\mu\nu}^{\phi\phi h}.
\label{eq:FlowEqDimensionful}
\end{align}
Note that the choice of singling out the classical tensor structure is an approximation, see also \Cref{app:VertexExpansion}. Nonetheless, the right-hand side of the flow equation \labelcref{eq:WetterichEquation} generates all tensor structures compatible with the symmetries. Then, in order to match it with the left-hand side, we have to define a projection scheme. While in the main text we focus on the classical projection, we report here one different projection scheme,
\begin{enumerate}
\item \emph{Classical projection}: The flow is being projected on the classical tensor structure alone, to wit
\begin{align}
\partial_t \bar{ \Gamma }^{\phi\phi h}_{\mu\nu} \mathcal{ T }^{ \phi\phi h, \mu\nu }\, .
\label{eq:classicalProjection}
\end{align}
This is the projection employed throughout this work.
\item \emph{Transverse traceless projection}: The flow is projected first with the transverse-traceless projection operator $\Pi_{tt}$, this isolates the physical graviton mode. Subsequently, we project by means of the classical structure
\begin{align}
            \partial_t \bar{ \Gamma }^{\phi\phi h}_{\mu\nu}\ \Pi_{tt}^{ \mu\nu\rho\sigma }\  \mathcal{ T }^{ \phi\phi h}_{ \rho\sigma }.
\label{eq:ttProjection}
\end{align}
\end{enumerate}
We refer to \Cref{app:projection of flow} for a comparison between the two schemes.

We employ classical projection \labelcref{eq:classicalProjection} on the flow in \labelcref{eq:FlowEqDimensionful}, and define the dimensionless coupling $g_{\phi\phi h}$,
\begin{align}
    G_{\phi\phi h} &= k^{-2}g_{\phi\phi h}\,, \notag \\[1ex]
    \partial_t G_{\phi\phi h} &= k^{-2}(\partial_t g_{\phi\phi h} - 2 g_{\phi\phi h})\,.
\label{eq:DimensionlessCouplings}
\end{align}
Thus, the flow equation reads
\begin{align}
    \partial_t g_{\phi\phi h} (p_i)&=
    \left( 2+ \sum_j \eta_j(p_j^2) \right) g_{\phi\phi h} (p_i)  \notag \\[1ex]
    &\quad\;+ 2g^{1/2}_{\phi\phi h} (p_i) \frac{ \partial_t \bar{\Gamma}^{\phi\phi h}_{\mu\nu} \mathcal{T}^{\phi\phi h, \mu\nu} }{ \mathcal{N} }\,,
\label{eq:FlowEqAppendix}
\end{align}
where we have defined the normalisation
\begin{align}
    \mathcal{N} &:= k^{-1}\, \mathcal{T}^{\phi\phi h, \mu\nu } \mathcal{T}^{\phi\phi h}_{\mu\nu} \notag \\[1ex]
 &= \frac{1}{2}\, p^4 (1 + z^2)k^{-1}\,.
\label{eq:NormClassicalExplicit}
\end{align}
The factor of $k^{-1}$ accounts for the mass dimension of the Newton coupling. Furthermore we introduce the shorthand notation $F_k$ for the flow part in \labelcref{eq:FlowEqAppendix},
\begin{align}
    F_k\!\left(p_i, \{ g_k^{(n)} \} \right) = 2\ \frac{ \partial_t \bar{\Gamma}^{\phi\phi h}_{\mu\nu} \mathcal{T}^{\phi\phi h, \mu\nu} }{ \mathcal{N} }\,.
\label{eq:ShorthandOfFlowAppendix}
\end{align}
Note that while the flow equation changes with the projection scheme, the structural form of the flow equation remains the same.

\section{Transverse-Traceless Projection}
\label{app:projection of flow}

We want to compare the classical projection \labelcref{eq:classicalProjection} with the transverse-traceless projection \labelcref{eq:ttProjection}. The former is used in the main text, and the corresponding flow is shown in \Cref{fig:FlowResolution}, while we show the flow of the latter in \Cref{fig:FlowResolutionTT}. The projected flows share many common features: both flows show a maximum around $z=-1/2$, which corresponds to the momentum-symmetric point. Furthermore, both flows are mostly negative, which leads to a positive fixed-point for the Newton coupling. In fact, the transverse-traceless flow is fully negative while the classical flow has some positive values for large $p/k$ and $z\to -1$. Overall, both flows are in qualitative agreement, which is precisely what one would hope for.

There is one surprising feature in the transverse-traceless flow: the flow diverges for $z\to1$, see \Cref{fig:FlowResolutionTT}. The reason is that the transverse-traceless projection has no overlap with the classical tensor structure in the limit $z\to 1$,
\begin{align}
\mathcal{ T }^{\phi\phi h}_{\mu\nu}\ \Pi_{tt}^{ \mu\nu\rho\sigma }\  \mathcal{ T }^{ \phi\phi h}_{ \rho\sigma }= \frac{1}{6} p^4 (z-1)^2\,,
\label{eq:TTnormalization}
\end{align}
while the corresponding flow is not vanishing
\begin{align}
 \lim_{z\to1 } \partial_t \Gamma^{\phi\phi h}_{\mu\nu}\ \Pi_{tt}^{ \mu\nu\rho\sigma }\  \mathcal{ T }^{ \phi\phi h}_{ \rho\sigma } \neq 0\,.
\end{align}
The reason is simply that the flow generates other tensor structures than the classical one, which are not resolved here, see \labelcref{eq:ScalarTensorBasis} for the full basis of the $\phi\phi h$ vertex. The issue is therefore resolved by utilising a complete set of tensor structures. Here, we instead use the classical projection in the main text.

\section{Locality of the Flow}
\label{app:localityFlow}
For UV momenta ($p/k\to\infty$), the function $F(p/k,z)$ freezes at a constant value, see \Cref{fig:FlowResolution,fig:FlowResolutionTT}. This implies that with momentum independent vertices, the flow is non-local \cite{Christiansen:2015rva}. The locality property implies that the integration of a shell around $k\simeq p_i$, being $p_i$ all the momentum transfers, does not influence the UV behaviour of the theory, i.e.~it is local in momentum space. This is formally encoded in the limit
\begin{align}
    \lim_{ p_i/k\to\infty } \frac{ \dot\Gamma^{(n)}_k(p_i) \circ \Pi }{ \Gamma_k^{(n)}(p_i) \circ \Pi } = 0\,,
\label{eq:localityProperty}
\end{align}
being $\Pi$ a suitable projection operator. For perturbatively renormalisable theories, the flow is local by power-counting, while the same is not guaranteed for perturbatively non-renormalisable theories, i.e.~quantum Einstein gravity. It is thus a non-trivial property that the flow of the transverse-traceless graviton two- and three-point functions is momentum local for Einstein-Hilbert tensor structures with constant vertex dressings, see \cite{Christiansen:2015rva, Christiansen:2014raa}, while this does not hold for the other modes of the two- and three-point function, see \cite{Reichert:2018nih, Knorr:2021niv}.

\begin{figure}[tbp]
    \includegraphics[width=0.9\columnwidth]{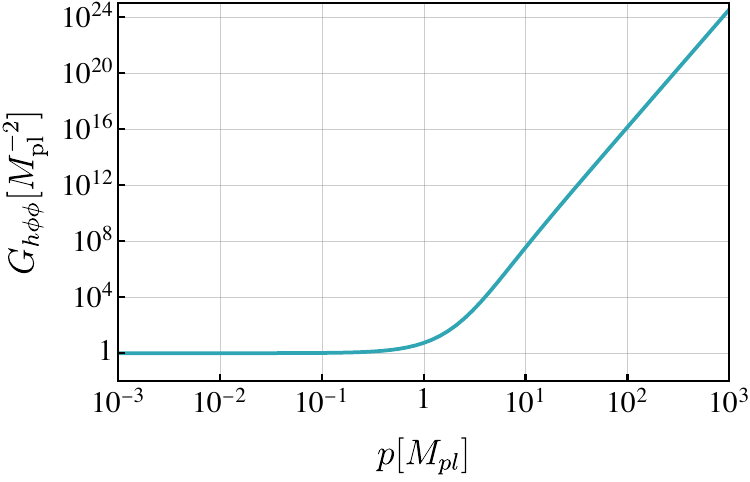}
    \caption{ Incorrect Newton coupling $G_{\phi\phi h}(p)$ obtained by solving \labelcref{eq:coarseApprox}. We point out the clear difference with the correct result \Cref{fig:FullAngularCoupling}. Such a difference is rooted in the improper treatment of the asymptotic behaviour of $F(p/k)$, see \Cref{fig:FlowResolution,fig:FlowResolutionTT}.}
    \label{fig:CoarseApproxSolution}
\end{figure}

To see how \labelcref{eq:localityProperty} translates to our setup, let us rewrite the vertex ansatz \labelcref{eq:VertexParametrization} as
\begin{align}
\Gamma^{\phi\phi h} (p_i) = \sqrt{ Z_{h,k}(p_h^2) } \sqrt{ G_{\phi\phi h,k}(p_i) }\ \mathcal{T}^{\phi\phi h}(p_i)\,,
\label{eq:vertexAnsatzRewritten}
\end{align}
where we set $Z_\phi=1$, see also \Cref{sec:ApproximationsAndConventions}. By projecting this according to \labelcref{eq:classicalProjection}, we obtain
\begin{align}
    \Gamma^{\phi\phi h}(p_i)\, \mathcal{T}^{\phi\phi h} = \sqrt{ Z_{h,k}(p_h^2) } \sqrt{ G_{\phi\phi h,k}(p_i) }\, \mathcal{N}\, k\,,
\end{align}
with the normalisation $\mathcal{N}$ as defined in \labelcref{eq:NormClassicalExplicit}. The ratio \labelcref{eq:localityProperty} then reads
\begin{align}
    \frac{  \partial_t \Gamma^{\phi\phi h}_k (p_i) \circ \Pi }{ \Gamma^{\phi\phi h}_k (p_i) \circ \Pi } &= 
    \frac{1}{g^{1/2}_{\phi\phi h,k}(p_i)}\, \frac{ \partial_t \bar{\Gamma}^{\phi\phi h}_k \, \mathcal{T}^{\phi\phi h} }{ \mathcal{N} } \notag  \\[1ex]
    &= g_{\phi\phi h, k}(p_i)\ F_k(p_i)\,,
\label{eq:LocalityOfFlowFunctionAndCoupling}
\end{align}
where we used \labelcref{eq:FlowAvatarIdentification} and the definition of dimensionless coupling. If we approximate \labelcref{eq:LocalityOfFlowFunctionAndCoupling} with a momentum-independent vertex dressing $g_{\phi\phi h.k}(0)$, then the limit \labelcref{eq:localityProperty} translates to
\begin{align}
    \lim_{p/k\to\infty} F_k(p_i) = 0\,,
\label{eq:localityOfF}
\end{align}
which is not fulfilled by \Cref{fig:FlowResolution,fig:FlowResolutionTT}. If instead we take into account the momentum dependence of the vertex dressing, given by the asymptotically safe scaling
\begin{align}
    g_{\phi\phi h,k}(p) \xrightarrow[]{p^2/k^2\gg1} \frac{k^2}{p^2}\,,
\label{eq:couplingScalingLocality}
\end{align}
then this suppresses \labelcref{eq:LocalityOfFlowFunctionAndCoupling} and ensures momentum locality \labelcref{eq:localityProperty} as long as $F_k(p)$ does not grow faster than $p^{-2}$, which is indeed the case.

In summary, momentum locality holds if the momentum dependence of the vertex dressing is taken into account. It is therefore necessary that a given approximation of the vertex running is able to take this momentum dependence into account, otherwise the result will display a wrong scaling behaviour in the UV, as we will show below. This is in contrast to flows where already a constant vertex dressing is sufficient to guarantee momentum locality, such as the flow of the transverse-traceless two- and three-point functions  \cite{Christiansen:2015rva, Christiansen:2014raa}.

We now show how neglecting momentum dependencies can lead to unphysical results at the example of the momentum-symmetric configuration $p=\|p_1\|=\|p_2\|$ and $z=-1/2$. We use an approximation that disentangles $g_{\phi\phi h}$ on the right-hand side of \labelcref{eq:FlowEquationFinalFinal} into a momentum-dependent and a momentum-independent part, i.e.,
\begin{align}
    \partial_t g_{\phi\phi h,k}(p) = &\left( 2+ \eta_h(p) \right) g_{\phi\phi h,k}(p) \notag \\[1ex]
    &+ g_{\phi\phi h,k}(p) g_{\phi\phi h,k}(p=0) F(p/k)\,.
\label{eq:coarseApprox}
\end{align}
\Cref{eq:coarseApprox} can be explicitly solved for $g_{\phi\phi h,k}(p)$, given an initial condition at some UV scale $k=\Lambda$. The explicit solution for the dimensionless coupling $g_{\phi\phi h}(p)$ reads
\begin{align}
    g_{\phi\phi h}(p) = g_\Lambda \exp\left[ \int_\Lambda^0 \frac{\mathrm{d} k}{k} \left( 2 + \eta_h(p) + g_{\phi\phi h,k}  F(p/k) \right) \right],
\label{eq:coarseApproxSolution}
\end{align}
where $g_\Lambda$ is the initial condition at the UV scale and $g_{\phi\phi h,k}$ is a trajectory at vanishing momentum. The result of this approximation for the dimensionful $G_{\phi\phi h}(p)$ is shown in \Cref{fig:CoarseApproxSolution}. The result is unphysical, as the Newton coupling does not allow for the correct scaling in the UV. Hence, a unjustified approximation can lead to unphysical results when the flow function with a constant vertex dressing does not satisfy \labelcref{eq:localityOfF}.

\begin{figure*}[tbp]
    \includegraphics[width=0.49\textwidth]{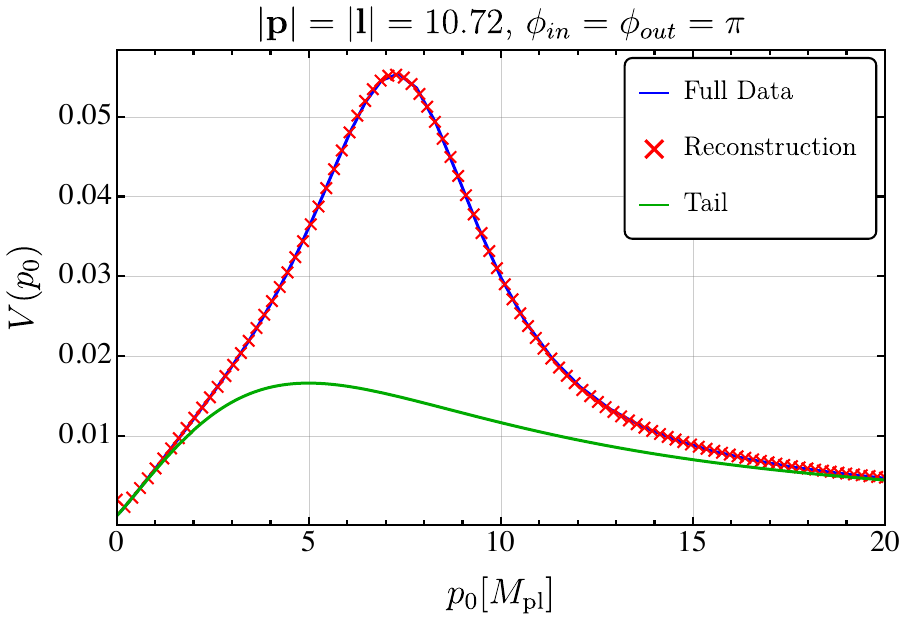}\hfill
    \includegraphics[width=0.49\textwidth]{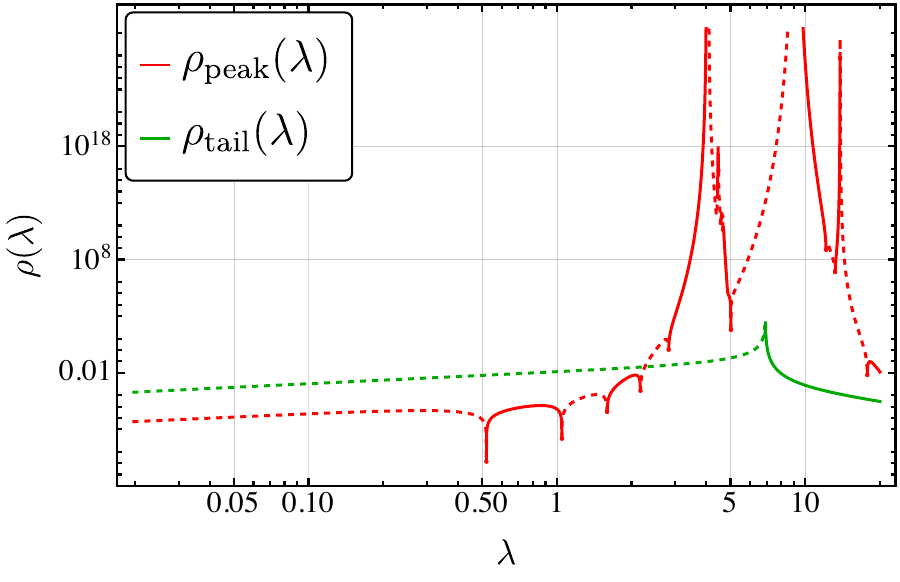}
    \caption{Reconstructed vertex (left) and spectral function (right) obtained via a Breit-Wigner fit for $\|\mathbf p\|=10.72$. The spectral function $\rho(\lambda)$ is highly unstable, probably due to overfitting.}
    \label{fig:bwReconstruction}
\end{figure*}

\section{Reconstruction Methods}
\label{app:comparisonRecMethods}

The reconstruction procedure is highly non-trivial due to the rich structure of the Euclidean vertex displayed in \Cref{fig:VertexExternalP}. While a full reconstruction of the vertex in the whole $p_0$ plane for all external momentum $\|\mathbf p\|$ would be beneficial, it is actually not necessary for our scope. After all, we are interested in the scattering amplitude and cross section, which can be computed by evaluating the vertex on-shell, i.e.~for $p_0 = \|\mathbf p\|$. In the main text, we resorted to a Padé-type reconstruction, which is not suitable for the whole range of $p_0$. However, for the on-shell value $p_0 = \|\mathbf p\|$, the Padé reconstruction is sufficiently accurate. Other methods that we have tested include a Breit-Wigner (BW) fit and a Gaussian Process Regression (GPR) reconstruction.

With the BW procedure, the main idea is to fit the Euclidean data with a sum and product of BW structures of the type
\begin{align}
   \text{BW}(p_0) = \left( \frac{1}{ (p_0 + \Gamma)^2 + M^2 } \right)^{\delta}.    
\label{eq:BWStructure}
\end{align}
By ensuring appropriate bounds on the parameters, one can guarantee that the upper-half of the complex plane of $p_0$ is free from cuts and poles, thus leading to a well-defined reconstruction. For further discussions on this, see \cite{Cyrol:2018xeq, Bonanno:2021squ}.

We separated the tails for $p_0\to0$ and $p_0\to\infty$ from the peak structure. For the asymptotic tails, we use a power law suppressed by a single BW,
\begin{align}
    V_\text{tail}(p_0) = a(p_0 + c)^b\ N_T \left( \frac{1}{ (p_0 + \Gamma_T)^2 + M_T^2} \right)^{\delta_T}.
\end{align}
The parameters $a,b,c$ are fixed by fitting the IR tail, while $N_T, M_T, \Gamma_T, \delta_T$ are fixed such that the UV decay $p_0^{-2}$ is reproduced for high momenta. Once the tails are known (green curve in \Cref{fig:bwReconstruction}), we subtract them from the data, and we fit the remaining peak structure with a sum of three products of five BWs,
\begin{align}
    V_\text{BW}(p_0) = \sum_{i=1}^3\ N_i\ \prod_{k=1}^{5}\ \left( \frac{1}{ ( p_0 + \Gamma_{i,k} )^2 + M_{i,k}^2 } \right)^{\delta_{i,k}}.
\end{align}
This ansatz (combined with the tail fit) leads to a sufficiently good reconstruction as shown in \Cref{fig:bwReconstruction} where we showcase the result for the external momentum $\|\mathbf p\|=10.72$. In turn, the spectral function associated with this is highly unstable, as shown on the right-hand side of \Cref{fig:bwReconstruction}. 

\begin{figure*}[tbp]
    \includegraphics[width= 0.47\textwidth]{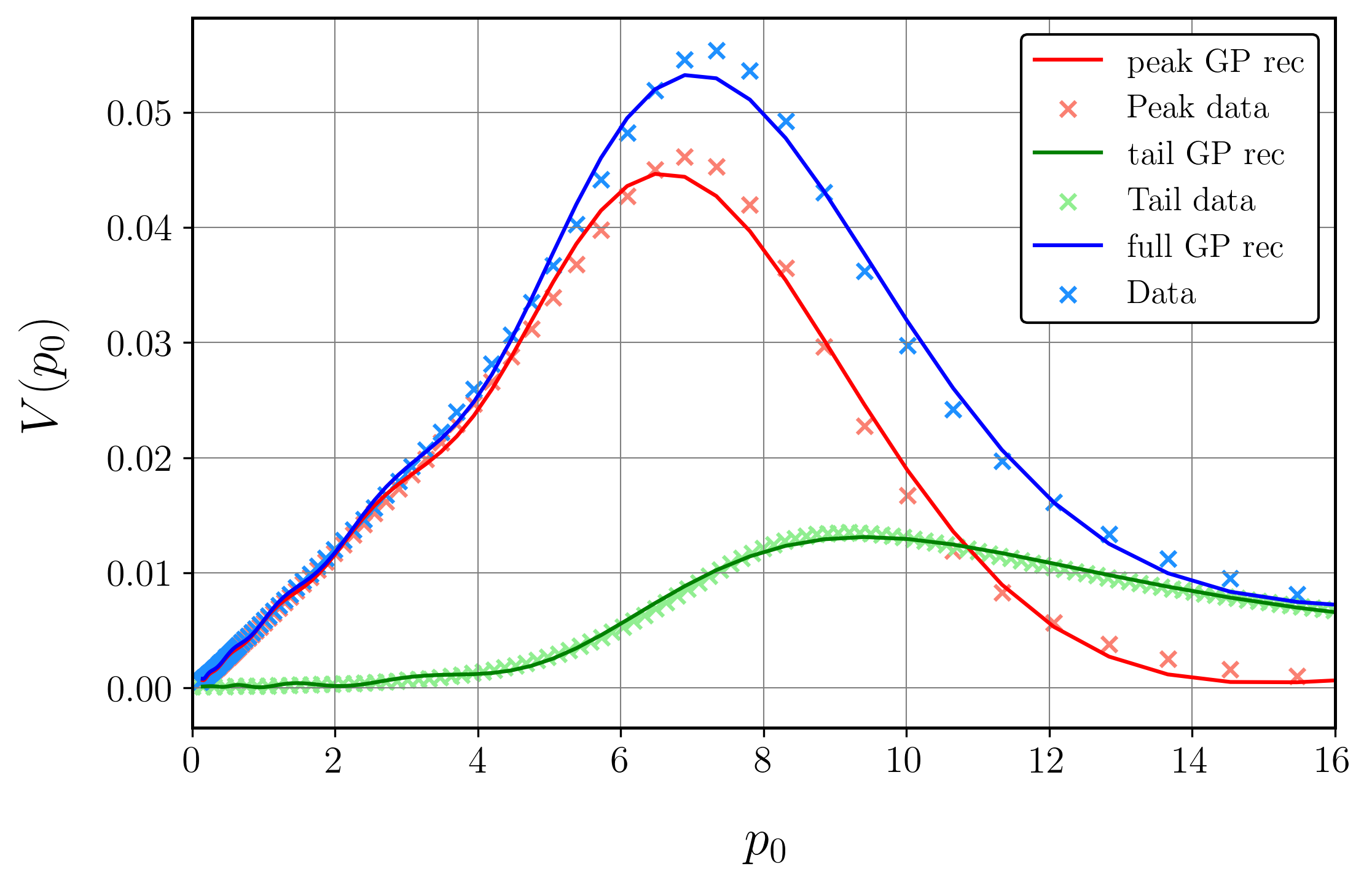}\hfill
    \includegraphics[width= 0.49\textwidth]{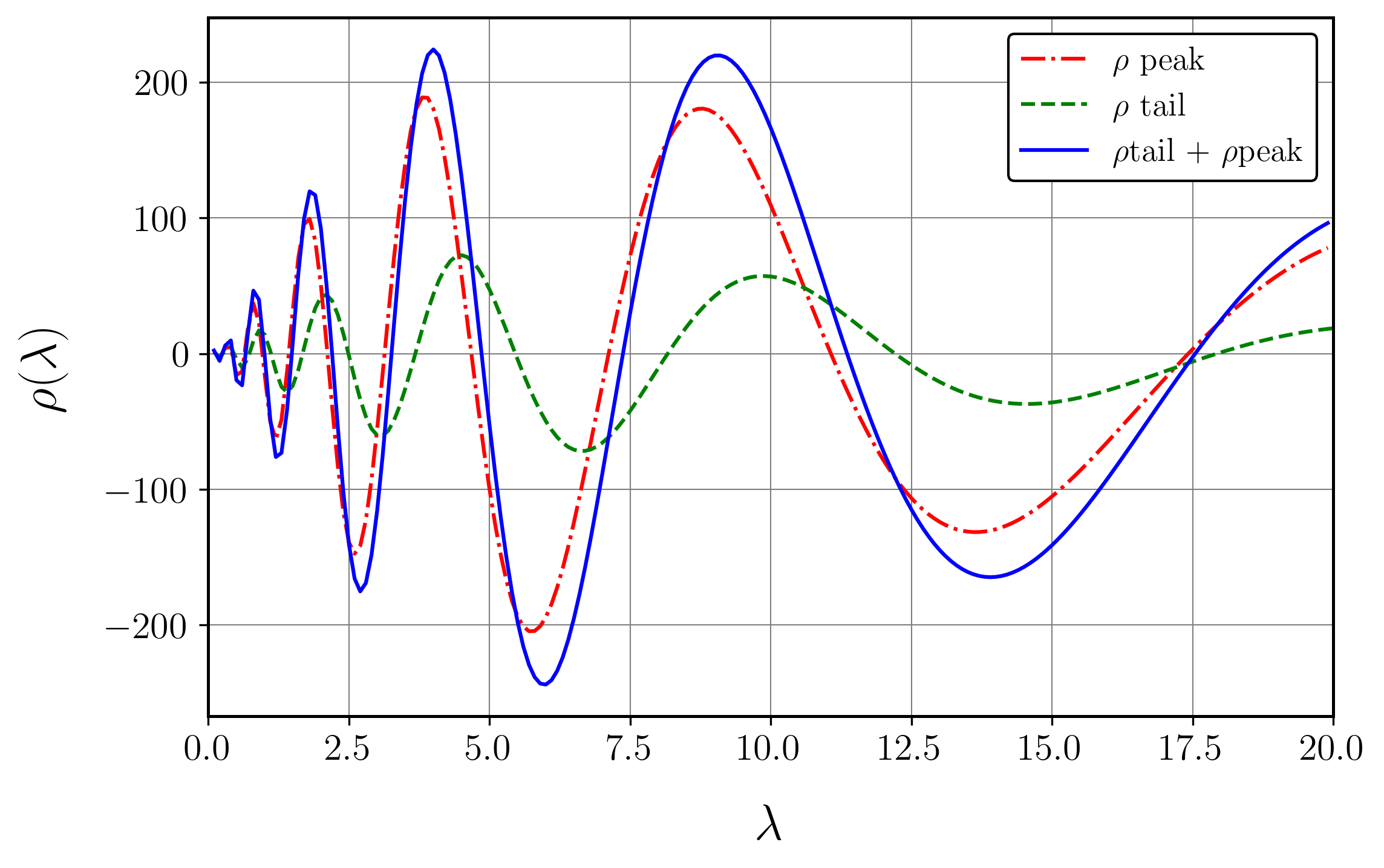}
    \caption{Reconstructed vertex (left) and spectral function (right) and obtained via GPR for $\|\mathbf p\|=10.72$. The spectral function is highly oscillatory.}
    \label{fig:GPRReconstructioncComparison}
\end{figure*}

We also tested Gaussian Process Regression (GPR), see \cite{Horak:2021syv} for details. The idea is to treat the data as a random variable with a given mean and covariance. The covariance is encoded in a kernel function, which is chosen according to the problem at hand, while the mean is usually set to zero. The kernel function is taken to be the radial basis function (RBF). The strength of this approach is the agnosticism on the functional form of the data, which is only assumed to be smooth. The result for both the reconstruction and the spectral function is shown in \Cref{fig:GPRReconstructioncComparison}. The reconstruction is comparable to the BW fit, but the spectral function also contains a lot of unphysical oscillations.

\subsection{Schlessinger Point Method}
\label{app:SPM}
The Schlessinger Point Method, also known as Resonances Via Pad{\'e} (RVP) method, is based on a rational-fraction representation similar to Pad{\'e} approximation methods. The rational-fraction construction interpolates a set of N points $(x_i, y_i)$ such that
\begin{align}
	C_{\text{N}} (x) = \dfrac{y_1}{1+\dfrac{a_1 (x - x_1)}{1+ \dfrac{a_2 (x - x_2)}{\vdots \dots a_{\text{N}-1} (x - x_{\text{N}-1})}}}  \, .
\end{align}
It is immediately apparent that $C_{\text{N}} (x_1) = y_1$ and that the coefficients $a_{1}, a_{2}, \ldots, a_{\text{N}-1}$ are chosen such that $C_{\text{N}} (x_i) = y_i \, \, \forall \, i$. They are determined by using a recursive formula that applies to every $a_i$ except $a_1$. For the latter, we have
\begin{align}
	a_1 = \frac{\left( y_1 / y_2 \right) - 1}{x_2 - x_1} \, ,
\end{align}
and in general
\begin{align}
	a_l = \frac{1}{x_l - x_{l+1}} \left\{ 1 + \dfrac{a_{l-1}(x_{l+1} - x_{l-1})}{1 + \dfrac{a_{l-2}(x_{l+1} - x_{l-2})}{\vdots \dots \dfrac{a_1 (x_{l+1} - x_1)}{1 - (y_1 / y_{l+1})}}} \right\} \, .
\end{align}

\newpage
\bibliography{bibliography}

\end{document}